\documentclass{article}
\usepackage{ijcai21}

\usepackage{times}
\usepackage{soul}
\usepackage{url}
\usepackage[hidelinks]{hyperref}
\usepackage[utf8]{inputenc}
\usepackage[small]{caption}
\usepackage{graphicx}
\usepackage{mathtools}
\usepackage{amssymb}
\usepackage{amsmath}
\usepackage{amsfonts}
\usepackage{amsthm}
\usepackage{booktabs}
\usepackage{algorithm}
\usepackage{algorithmic}
\usepackage{xcolor}

\usepackage{algorithm}
\usepackage{algorithmic}
\usepackage[algo2e, noend, noline, ruled, linesnumbered]{algorithm2e}

\usepackage{caption}
\usepackage{subcaption}
\usepackage{selectp}

\newcommand{\EE}{\mathbb{E}}
\newcommand{\RR}{\mathbb{R}}
\newcommand{\cA}{\mathcal{A}}
\newcommand{\cP}{\mathcal{P}}
\newcommand{\cS}{\mathcal{S}}
\newcommand{\algo}{Flock'n RL\xspace}
\DeclareMathOperator*{\argmax}{arg\,max}

\title{Mean Field Games Flock! The Reinforcement Learning Way}

\author{
Sarah Perrin$^1$\and
Mathieu Laurière$^2$\and
Julien Pérolat$^3$\and
Matthieu Geist$^4$ \and
Romuald Élie$^3$ \And
Olivier Pietquin$^4$ 
\affiliations
$^1$Univ. Lille, CNRS, Inria, Centrale Lille, UMR 9189 CRIStAL\\
$^2$Princeton University, ORFE\\
$^3$DeepMind Paris\\
$^4$Google Research, Brain Team
\emails
sarah.perrin@inria.fr,
lauriere@princeton.edu,
\{perolat, mfgeist, relie, pietquin\}@google.com,
}

\begin{document}

\maketitle

\begin{abstract}
  We present a method enabling a large number of agents to learn how to flock, which is a natural behavior observed in large populations of animals. 
  This problem has drawn a lot of interest but requires many structural assumptions and is tractable only in small dimensions. We phrase this problem as a Mean Field Game (MFG), where each individual chooses its acceleration depending on the population behavior. Combining Deep Reinforcement Learning (RL) and Normalizing Flows (NF), we obtain a tractable solution requiring only very weak assumptions.
  Our algorithm finds a Nash Equilibrium and the agents adapt their velocity to match the neighboring flock's average one. 
  We use Fictitious Play and alternate: (1) computing an approximate best response with Deep RL, and (2) estimating the next population distribution with NF. We show numerically that our algorithm learn multi-group or high-dimensional flocking with obstacles. 
\end{abstract}

\section{Introduction}

The term flocking describes the behavior of large populations of birds that fly in compact groups with similar velocities, often exhibiting elegant motion patterns in the sky. Such behavior  
is pervasive in the animal realm, from fish to birds, bees or ants. This intriguing property has been widely studied in the scientific literature~\cite{shaw1975naturalist,reynolds1987flocks} 
and its modeling finds applications in psychology, 
animation, 
social science, 
or swarm robotics. One of the most popular approaches to model flocking was proposed in~\cite{cucker2007emergent} and allows predicting the evolution of each agent's velocity from the speeds of its neighbors. 

To go beyond pure description of population behaviours and emphasize on the decentralized aspect of the underlying decision making process, this model has been revisited to integrate an optimal control perspective, see \textit{e.g.}~\cite{caponigro2013sparse,bailo2018optimal}. 
Each agent controls its velocity and hence its position by dynamically adapting its acceleration so as to maximize a reward that depends on the others' behavior. An important question is a proper understanding of the nature of the equilibrium reached by the population of agents, emphasizing how a consensus can be reached in a group without centralized decisions. Such question is often studied using the notion of Nash equilibrium and becomes extremely complex when the number of agents grows.

A way to approximate Nash equilibria in large games is to study the limit case of an continuum of identical agents, in which the local effect of each agent becomes negligible. This is the basis of the Mean Field Games (MFGs) paradigm introduced in~\cite{MR2295621}. MFGs have found numerous applications from economics 
to energy production 
and engineering. 
A canonical example is crowd motion modeling  
in which pedestrians want to move while avoiding congestion effects. In flocking, the purpose is different since the agents intend to remain together as a group, but the mean-field approximation can still be used to mitigate the complexity. 

However, finding an equilibrium in MFGs is computationally intractable when the state space exceeds a few dimensions. In traditional flocking models, each agent's state is described by a position and a velocity, while the control is the acceleration. 
In terms of computational cost, this typically rules out state-of-the-art numerical techniques for MFGs based on finite difference schemes for partial differential equations (PDEs)~\cite{MR2679575}. In addition, PDEs are in general hard to solve when the geometry is complex and require full knowledge of the model.

 For these reasons, Reinforcement Learning (RL)
 to learn control strategies for MFGs has recently gained in popularity~\cite{guo2019learning,elie2020convergence,perrin2020fictitious}. Combined with deep neural nets, RL has been used successfully to tackle problems which are too complex to be solved by exact methods~\cite{Silver18AlphaZero} or to address learning in multi-agent systems~\cite{Lanctot17PSRO}.  Particularly relevant in our context, are works providing techniques to compute an optimal policy~\cite{SACDBLP:journals/corr/abs-1801-01290,lillicrap2015continuous} 
 and methods to approximate probability distributions in high dimension~\cite{rezende2015variational,kobyzev2020normalizing}.

Our \textbf{main contributions} are: \textbf{(1)} we cast the flocking problem into a MFG and propose variations which allow multi-group flocking as well as flocking in high dimension with complex topologies, \textbf{(2)} we introduce the \algo algorithm that builds upon the Fictitious Play paradigm and involves deep neural networks and RL to solve the model-free flocking MFG, and \textbf{(3)} we illustrate our approach on several numerical examples and evaluate the solution with approximate performance matrix and exploitability.

\section{Background}

Three main formalisms will be combined: flocking, Mean Field Games (MFG), Reinforcement Learning (RL).

$\EE_z$ stands for the expectation w.r.t. the random variable $z$.

\subsection{The model of flocking} 
\label{sec:background-Flocking}

To model a flocking behaviour, we consider the following system of $N$ agents derived in a discrete time setting in \cite{nourian2011mean} from Cucker-Smale flocking modeling. Each agent $i$ has a position and a velocity, each in dimension $d$ and denoted respectively by $x^i$ and $v^i$. We assume that it can control its velocity by choosing the acceleration, denoted by $u^i$. The dynamics of agent $i \in \left\{1, \dots, N \right\}$ is:

$$
        x^i_{t+1} = x^i_{t} + v^i_{t}\Delta t, \qquad
        v^i_{t+1} = v^i_{t} + u^i_{t}\Delta t + \epsilon^i_{t+1},
$$
where $\Delta t$ is the time step and  $\epsilon^i_{t}$ is a random variable playing the role of a random disturbance. 
We assume that each agent is optimizing for a \textit{flocking criterion} $f$ that is underlying to the flocking behaviour. For agent $i$ at time $t$, $f$ is of the form:
\begin{equation}
    \label{eq_reward_n_agents_generic}
    f^i_{t} = f(x^i_{t}, v^i_{t}, u^i_{t}, \mu^N_t),
\end{equation}
where the interactions with other agents are only through the empirical distribution of states and velocities denoted by:
$
    \mu^N_t = \frac{1}{N} \sum_{j=1}^N \delta_{(x^j_t, v^j_t)}.
$

We focus on criteria incorporating a term of the form:
\begin{equation}
\footnotesize
\label{eq:def-f-flock-MF}
    f^{\mathrm{flock}}_{\beta}(x,v,u,\mu) 
    =
    \text{-} \left\| \int_{\RR^{2d}}  \frac{\left(v - v' \right)}{ (1 + \left\| x - x' \right\|^2 )^\beta} \,d\mu(x', v')  \right\|^2 ,
\end{equation}
where $\beta \ge 0$ is a parameter and $(x,v,u, \mu) \in \RR^d \times \RR^d \times \RR^d \times \mathcal{P}(\RR^d \times \RR^d)$, with $\mathcal{P}(E)$ denoting the set of probability measures on a set $E$. This criterion incentivises agents to align their velocities, especially if they are close to each other. Note that $\beta$ parameterizes the level of interactions between agents and strongly impacts the flocking behavior: if $\beta=0$, each agent tries to align its velocity with all the other agents of the population irrespective of their positions, whereas the larger $\beta>0$, the more importance is given to its closest neighbors (in terms of position).

In the $N$-agent case, for agent $i$, it becomes:
\begin{equation}
    f^{\mathrm{flock},i}_{\beta,t} = - \left\| \frac{1}{N} \sum^N_{j=1} \frac{(v^i_{t} - v^j_{t} )}{ (1 + \| x^i_{t} - x^j_{t} \|^2 )^\beta}  \right\|^2.
    \label{eq_reward_n_agents_flocking}
\end{equation}
 The actual criterion will typically include other terms, for instance to discourage agents from using a very large acceleration, or to encourage them to be close to a specific position. We provide such examples in Sec.~\ref{sec:expe}.

Since the agents may be considered as selfish (they try to maximize their own criterion) and may have conflicting goals (\textit{e.g.} different desired velocities), we consider Nash equilibrium as a notion of solution to this problem and the individual criterion can be seen as the payoff for each agent. The total payoff of agent $i$ given the other agents' strategies $u^{-i} = (u^1,\dots,u^{i-1},u^{i+1}, \dots, u^N)$ is:
$
    F^i_{u^{-i}}(u^i) = \EE_{x^i_{t},v^i_{t}}\Big[\sum_{t \ge 0} \gamma^t  f^i_{t} \Big],
$
with $f^i_{t}$ defined  Eq.~\eqref{eq_reward_n_agents_generic}.  
In this context, a \emph{Nash equilibrium} is  a strategy profile $(\hat u^1, \hat u^2,\dots, \hat u^N)$ such that there's no profitable unilateral deviation, \textit{i.e.}, for every $i=1,\dots,N$, for every control $u^i$,
$
    F^i_{\hat u^{-i}}(\hat u^i) \ge F^i_{\hat u^{-i}}(u^i).
$

\subsection{Mean Field Games}
\label{sec:background-MFG}

An MFG describes a game for a continuum of identical agents and is fully characterized by the dynamics and the \textit{payoff} function of a representative agent. More precisely, denoting by $\mu_t$ the state distribution of the population, and by $\xi_t \in \RR^\ell$ and $\alpha_t \in \RR^k$ the state and the control of an infinitesimal agent, the dynamics of the infinitesimal agent is given by
\begin{equation}
    \label{eq:dyn-x-MFG}
    \xi_{t+1} = \xi_{t} + b(\xi_t, \alpha_t,\mu_t) + \sigma \epsilon_{t+1},
\end{equation}
where $b: \RR^\ell \times \RR^k \times \mathcal{P}(\RR^\ell) \to \RR^\ell$ is a drift (or transition) function, $\sigma$ is a $\ell \times \ell$ matrix and $\epsilon_{t+1}$ is a noise term taking values in $\RR^\ell$. We assume that the sequence of noises $(\epsilon_t)_{t\geq 0}$ is i.i.d. (\textit{e.g.} Gaussian). The objective of each infinitesimal agent is to maximize its total expected payoff, defined given a flow of distributions $\mu = (\mu_t)_{t \ge 0}$ and a strategy $\alpha$ (\textit{i.e.}, a stochastic process adapted to the filtration generated by $(\epsilon_t)_{t \ge 0}$) as:
$
    J_\mu(\alpha) = \EE_{\xi_t,\alpha_t}\Big[ \sum_{t \ge 0} \gamma^t \varphi(\xi_t, \alpha_t,\mu_t)\Big],
$
where $\gamma \in (0,1)$ is a discount factor and $\varphi: \RR^\ell \times \RR^k \times \mathcal{P}(\RR^\ell) \to \RR^\ell$ is an instantaneous payoff function. Since this payoff depends on the population's state distribution, and since the other agents would also aim to maximize their payoff, a natural approach is to generalize the notion of Nash equilibrium to this framework. A \emph{mean field (Nash) equilibrium} is defined as a pair $(\hat\mu,\hat\alpha) = (\hat\mu_t,\hat\alpha_t)_{t \ge 0}$ of a flow of distributions and strategies such that the following two conditions are satisfied: $\hat\alpha$ is a best response against $\hat\mu$ (optimality) and $\hat\mu$ is the distribution generated by $\hat\alpha$ (consistency), \textit{i.e.},
\begin{enumerate}
    \item $\hat\alpha$ maximizes $\alpha \mapsto J_{\hat\mu}(\alpha)$;
    \item for every $t \ge 0$, $\hat\mu_t$ is the distribution of $\xi_t$ when it follows the dynamics~\eqref{eq:dyn-x-MFG} with $(\alpha_t,\mu_t)$ replaced by $(\hat\alpha_t,\hat\mu_t)$.
\end{enumerate}
Finding a mean field equilibrium thus amounts to finding a fixed point in the space of (flows of) probability distributions. The existence of equilibria can be proven through classical fixed point theorems \cite{CarmonaDelarue_book_I}. In most mean field games considered in the literature, the equilibrium is unique, which can be proved using either a strict contraction argument or the so-called Lasry-Lions monotonicity condition~\cite{MR2295621}. Computing solutions to MFGs is a challenging task, even when the state is in small dimension, due to the coupling between the optimality and the consistency conditions. This coupling typically implies that one needs to solve a forward-backward system where the forward equation describes the evolution of the distribution and the backward equation characterizes the optimal control. One can not be solved prior to the other one, which leads to numerical difficulties. The basic approach, which consists in iteratively solving each equation, works only in very restrictive settings and is otherwise subject to cycles. A method which does not suffer from this limitation is Fictitious Play, summarized in Alg.~\ref{algo:FP-MFG-general}. It consists in computing the best response against a weighted average of past distributions instead of just the last distribution. This algorithm has been shown to converge for more general MFGs. State-of-the-art numerical methods for MFGs based on partial differential equations can solve such problems with a high precision when the state is in small dimension and the geometry is  elementary~\cite{MR2679575,MR3148086}. 
More recently, numerical methods based on machine learning tools have been developed~\cite{CarmonaLauriere_DL,ruthotto2020machine}.  
These techniques rely on the full knowledge of the model and are restricted to classes of quite simple MFGs.
\begin{algorithm2e}[t!]
\SetKwInOut{Input}{input}\SetKwInOut{Output}{output}
\Input{ MFG = $\{\xi, \varphi, \mu_0\}$; \# of iterations $J$ ;} 
\caption{Generic Fictitious Play in MFGs \label{algo:FP-MFG-general}}
Define $\bar \mu_0 = \mu_0$ 
\For{$j=1,\dots,J$}{
  \label{alg:FP-BR-general} \hspace{-0.25cm}{\bf 1.} Set best response $\alpha_j = $ $\argmax_{\alpha} J_{\bar{\mu}_{j-1}} (\alpha)$ and let $\bar \alpha_j$ be the average of $(\alpha_i)_{i=0,\dots,j}$ 
  \label{alg:FP-distrib-general} \\
  \hspace{-0.25cm}{\bf 2.} Set $\mu_j = $  $\gamma$-stationary distribution induced by $\alpha_j$ \\
  \hspace{-0.25cm}{\bf 3.} Set $\bar \mu_j  = \frac{j-1}{j}\bar \mu_{j-1} + \frac{1}{j}\mu_j$ 
  \label{alg:FP-barmu-general}
}
\Return{$\bar \alpha_J, \bar \mu_J$}
\end{algorithm2e}
\subsection{Reinforcement Learning } 

\label{sec:background-RL}

The Reinforcement Learning (RL) paradigm 
is the machine learning answer to the optimal control problem. It aims at learning an optimal policy for an agent that interacts in an environment composed of states, by performing actions. Formally, the problem is framed under the Markov Decision Processes (MDP) framework. 
 An MDP is a tuple $(\cS, \cA, p, r, \gamma)$ where $\cS$ is a state space, $\cA$ is an action space, $p: \cS \times \cA \to \cP(\cS)$ is a transition kernel, $r:\cS \times \cA \to \RR $ is a \textit{reward} function and $\gamma$ is a discount factor (see Eq.~\eqref{eq:def-total-reward-MDP}). Using action $a$ when the current state is $s$ leads to a new state distributed according to $P(s,a)$ and produces a reward  $R(s,a)$. A policy $\pi: \cS \to \cP(\cA)$, $s \mapsto \pi(\cdot | s)$ provides a distribution over actions for each state. RL aims at learning a policy $\pi^*$ which maximizes the total return defined as the expected (discounted) sum of future  rewards:
\begin{equation}
    \label{eq:def-total-reward-MDP}
    R(\pi) = \EE_{a_t, s_{t+1}}\Big[\sum_{t \ge 0} \gamma^t  r(s_t,a_t) \Big],
\end{equation}
with $a_t \sim \pi(\cdot|s_t)$ and $s_{t+1} \sim p(\cdot|s_t, a_t)$. Note that if the dynamics ($p$ and $r$) is known to the agent, the problem can be solved using \textit{e.g.} dynamic programming. Most of the time, these quantities are unknown and RL is required. A plethora of algorithms exist to address the RL problem. 
Yet, we need to focus on methods that allow continuous action spaces as we want to control accelerations. One category of such algorithms is based on the Policy Gradient (PG) theorem~\cite{sutton:nips12} and makes use of the gradient ascent principle: 
$\pi \leftarrow \pi + \alpha \frac{\partial R(\pi) }{\partial \pi},$
where $\alpha$ is a learning rate. Yet, PG methods are known to be high-variance because they use Monte Carlo rollouts to estimate the gradient. A vast literature thus addresses the variance reduction problem. Most of the time, it involves an hybrid architecture, namely Actor-Critic, which relies on both a representation of the policy and of the so-called state-action value function $(s,a) \mapsto Q^\pi(s,a)$. $Q^\pi(s,a)$ is the total return conditioned on starting in state $s$ and using action $a$ before using policy $\pi$ for subsequent time steps. It can be estimated by bootstrapping, using the Markov property, through the Bellman equations. Most recent implementations rely on deep neural networks to approximate $\pi$ and $Q$ (\textit{e.g.} \cite{SACDBLP:journals/corr/abs-1801-01290}).

\section{Our approach}

In this section, we put together the pieces of the puzzle to numerically solve the flocking model. Based on a mean-field approximation, we first recast the flocking model as an MFG with decentralized decision making, for which we propose a numerical method relying on RL and deep neural networks.

\subsection{Flocking as an MFG}

\textbf{Mean field limit. } We go back to the model of flocking introduced in Sec.~\ref{sec:background-Flocking}. When the number of agents grows to infinity, the empirical distribution $\mu^N_t$ is expected to converge towards the law $\mu_t$ of $(x_t, v_t)$, which represents the position and velocity of an infinitesimal agent and have dynamics:
$$
        x_{t+1} = x_t + v_t\Delta t, \qquad
        v_{t+1} = v_t + u_t\Delta t + \epsilon_{t+1}.
$$
This problem can be viewed as an instance of the MFG framework discussed in Sec.~\ref{sec:background-MFG}, by taking the state to be the position-velocity pair and the action to be the acceleration, \textit{i.e.}, the dimensions are $\ell = 2d$, $k=d$, and $\xi = (x,v)$, $\alpha = u$. To accommodate for the degeneracy of the noise as only the velocities are disturbed, we take $\sigma = \begin{pmatrix} 0_d & 0_d \\ 0_d & 1_d \end{pmatrix}$ where $1_d$ is the $d$-dimensional identity matrix.

The counterpart of the notion of $N$-agent Nash equilibrium is an MFG equilibrium as described in Sec.~\ref{sec:background-MFG}. We focus here on equilibria which are stationary in time. In other words, the goal is to find a pair $(\hat \mu,\hat u)$ where $\hat \mu \in \mathcal{P}(\RR^\ell \times \RR^\ell)$ is a position-velocity distribution and $\hat u:\RR^\ell \times \RR^\ell \mapsto \RR^\ell$ is a feedback function to determine the acceleration given the position and velocity, such that: {\bf (1)} $\hat \mu$ is an invariant position-velocity distribution if the whole population uses the acceleration given by $\hat u$, and {\bf (2)} $\hat u$ maximizes the rewards when the agent's initial position-velocity distribution is $\hat\mu$ and the population distribution is $\hat \mu$ at every time step. In mathematical terms, $\hat u$ maximizes
$
    J_{\hat\mu}(u) = \EE_{x_t, v_t, u_t}\Big[ \sum_{t \ge 0} \gamma^t \varphi(x_t, v_t, u_t, \hat\mu)\Big],
$
where $(x_t,v_t)_{t \ge 0}$ is the trajectory of an infinitesimal agent who starts with distribution $\hat\mu$ at time $t=0$ and is controlled by the acceleration $(u_t)_{t \ge 0}$. As the payoff function $\varphi$ we use $f^{\mathrm{flock}}_{\beta}$ from Eq.~\eqref{eq:def-f-flock-MF}. Moreover, the consistency condition rewrites as: $\hat\mu$ is the stationary distribution of $(x_t,v_t)_{t \ge 0}$ if controlled by $(\hat u_t)_{t \ge 0}$.

\textbf{Theoretical analysis. }
The analysis of MFG with flocking effects is challenging due to the unusual structure of the dynamics 
and the payoff, which encourages gathering of the population. This is running counter to the classical Lasry-Lions monotonicity condition~\cite{MR2295621}, which typically penalizes the agents for being too close to each other. However, existence and uniqueness have been proved in some cases. If $\beta = 0$, every agent has the same influence over the representative agent and it is possible to show that the problem reduces to a Linear-Quadratic setting. Th~2 in \cite{nourian2011mean} shows that a mean-consensus in velocity is reached asymptotically with individual asymptotic variance $\frac{\sigma^2}{2}$.  If $\beta > 0$, \cite{nourian2011mean} shows that if the MF problem admits a unique solution, then there exists an $\epsilon_N$ Nash equilibrium for the $N$ agents problem and $\lim\limits_{N \rightarrow +\infty} \epsilon_N = 0$. Existence has also been proved when $\beta>0$ in \cite[Section 4.7.3]{CarmonaDelarue_book_I}, with a slight modification of the payoff, namely considering, with $\varphi$ a bounded function,   
    $r_t = -\varphi\Big(\Big\| \int_{\mathbb{R}^{2d}} \frac{\left(v_t - v' \right)}{ \left(1 + \left\| x_t - x' \right\|^2 \right)^\beta} \mu_t(dx',dv')  \Big\|^2\Big)$.

\subsection{The \algo Algorithm}

We propose Flock'n RL, a deep RL version of the Fictitious Play algorithm for MFGs~\cite{elie2020convergence} adapted to flocking. We consider a $\gamma$-discounted setting with continuous state and action spaces and we adapt Alg.~\ref{algFP-FlocknRL} from its original tabular formulation~\cite{perrin2020fictitious}. It alternates 3 steps: (1) estimation of a best response (using Deep RL) against the mean distribution  $\bar \mu_{j-1}$, which is fixed during the process, (2) estimation (with Normalizing Flows~\cite{kobyzev2020normalizing}) of the new induced distribution from trajectories generated by the previous policy, and (3) update of the mean distribution $\bar \mu_j$.


\begin{algorithm2e}[t!]
\SetKwInOut{Input}{input}\SetKwInOut{Output}{output}
\Input{  MFG = $\{(x,v), f^{\mathrm{flock}}_{\beta} , \mu_0\}$; \# of iterations $J$}
\caption{\algo \label{algFP-FlocknRL}}
Define $\bar \mu_0 = \mu_0$ 
\For{$j=1,\dots,J$}{
  \label{alg:FP-BR} 
  \hspace{-0.25cm}{\bf 1.} 
  Set best response $\displaystyle\pi_j = \argmax_{\pi} J_{\bar{\mu}_{j-1}} (\pi)$ with \textbf{SAC} and let $\bar \pi_j$ be the average of $(\pi_0, \dots, \pi_j)$
  \label{alg:FP-distrib} \\
  \hspace{-0.25cm}{\bf 2.} Using a \textbf{Normalizing Flow}, compute $\mu_j = $   $\gamma$-stationary distribution induced by $\pi_j$ \\
  \hspace{-0.25cm}{\bf 3.} Using a \textbf{Normalizing Flow} and samples from $(\mu_0, \dots, \mu_{j-1})$, estimate $\bar \mu_j$
  \label{alg:FP-barmu}
}
\Return{$\bar \pi_J, \bar \mu_J$}
\end{algorithm2e}

\subsubsection{Computing the Best Response with SAC}

The first step in the loop of Alg.~\ref{algFP-FlocknRL} is the computation of a best response against $\bar\mu_j$. In fact, the problem boils down to solving an MDP in which $\bar\mu_j$ enters as a parameter. Following the notation introduced in Sec.~\ref{sec:background-RL}, we take the state and action spaces to be respectively $\cS = \RR^{2d}$ (for position-velocity pairs) and $\cA = \RR^d$ (for accelerations). Letting $s = (x, v)$ and $a = u$, the reward is:
$
    (x,v,u) = (s,a) \mapsto r(s, a) = f(x, v, u, \bar\mu_j),
$
which depends on the given distribution $\bar\mu_j$. Remember that $r$ is the reward function of the MDP while $f$ is the optimization criterion in the flocking model.

As we set ourselves in continuous state and action spaces and in possibly high dimensions, we need an algorithm that scales. We choose to use Soft Actor Critic (SAC) \cite{SACDBLP:journals/corr/abs-1801-01290}, an off-policy actor-critic deep RL algorithm using entropy regularization. SAC is trained to maximize a trade-off between expected return and entropy, which allows to keep enough exploration during the training. It is designed to work on continuous action spaces, which makes it suited for acceleration controlled problems such as flocking.

The best response is computed against $\bar \mu_j$, the fixed average distribution at step $j$ of \algo. SAC maximizes the reward which is a variant of $f^{\mathrm{flock},i}_{\beta,t}$ from Eq.~\eqref{eq_reward_n_agents_flocking}. It needs samples from $\bar \mu_j$ in order to compute the positions and velocities of the fixed population. Note that, in order to measure more easily the progress during the learning at step $j$, we sample $N$ agents from $\bar \mu_j$ at the beginning of step 1 (\textit{i.e.} we do not sample new agents from $\bar \mu_j$ every time we need to compute the reward). During the learning, at the beginning of each episode, we sample a starting state $s_0 \sim \bar \mu_j$.

In the experiments, we will not need $\bar\pi_j$ but only the associated reward (see the exploitability metric in Sec.~\ref{sec:expe}). To this end, it is enough to keep in memory the past policies $(\pi_0, \dots, \pi_j)$ and simply average the induced rewards.

\subsubsection{Normalizing Flow for Distribution Embedding}

We choose to represent the different distributions using a generative model because the continuous state space prevents us from using a tabular representation. Furthermore, even if we could choose to discretize the state space, we would need a huge amount of data points to estimate the distribution using methods such as kernel density estimators. 
In dimension 6 (which is the dimension of our state space with 3-dimensional positions and velocities), such methods already suffer from the curse of dimensionality. 

Thus, we choose to estimate the second step  of Alg.~\ref{algFP-FlocknRL} using a Normalizing Flow (NF)~\cite{rezende2015variational,kobyzev2020normalizing}, which is a type of generative model, different from  Generative Adversarial Networks (GAN) 
or Variational Autoencoders (VAE). 
A flow-based generative model is constructed by a sequence of invertible transformations and allows efficient sampling and distribution approximation. Unlike GANs and VAEs, the model explicitly learns the data distribution and therefore the loss function simply identifies to the negative log-likelihood. An NF transforms a simple distribution (\textit{e.g. }Gaussian) into a complex one by applying a sequence of invertible transformations. In particular, a single transformation function $f$ of noise $z$ can be written as $x = f(z)$ where $z \sim h(z)$. Here, $h(z)$ is the noise distribution and will often be in practice a normal distribution.

Using the change of variable theorem, the probability density of $x$ under the flow can be written as: 
$
    p(x) = h(f^{-1}(x)) \left| \text{det}\left( \frac{\partial f^{-1}}{\partial x}\right) \right| .
$
We thus obtain the probability distribution of the final target variable. In practice, the transformations $f$ and $f^{-1}$ can be approximated by neural networks. 
Thus, given a dataset of observations (in our case rollouts from the current best response), the flow is trained by maximizing the total log likelihood $\sum_n \log p(x^{(n)})$.

\subsubsection{Computation of $\bar \mu_j$} 
Due to the above discussion on the difficulty to represent the distribution in continuous space and high dimension, the third step (Line~\ref{alg:FP-barmu} of Alg.~\ref{algFP-FlocknRL}) can not be implemented easily. We represent every $\mu_j$ as a generative model, so we can not ``average'' the normalizing flows corresponding to $(\mu_i)_{i=1,\dots,j}$ in a straightforward way but we can sample data points $x \sim \mu_i$ for each $i=1,\dots,j$. To have access to $\bar \mu_j$, we keep in memory every model $\mu_j$, $j \in \left\{1, \dots, J \right\}$ and, in order to sample points according to $\bar\mu_j$ for a fixed $j$, we sample points from $\mu_i, i \in \{1,\dots,j\},$ with probability $1/j$. These points are then used to learn the distribution $\bar\mu_j$ with an NF, as it is needed both for the reward and to sample the starting state of an agent during the process of learning a best response policy.

\section{Experiments}
\label{sec:expe}

\noindent 
\textbf{Environment. } We implemented the environment as a custom OpenAI gym environment 
to benefit from  the powerful gym framework and use the algorithms available in stable baselines~\cite{stable-baselines}. We define a state $s \in S$ as $s=(x,v)$ where $x$ and $v$ are respectively the vectors of positions and velocities. Each coordinate $x_i$ of the position can take any continuous value in the $d$-dimensional box $x_i \in [-100, +100]$, while the velocities are also continuous and clipped $v_i \in [-1,1]$. The state space for the positions is a torus, meaning that an agent reaching the box limit reappears at the other side of the box. We chose this setting to allow the agents to perfectly align their velocities (except for the effect of the noise), as we look for a stationary solution.

At the beginning of each iteration $j$ of Fictitious Play, we initialize a new gym environment with the current mean distribution $\bar \mu_j$, in order to compute the best response. 

\textbf{Model - Normalizing Flows. } To model distributions, we use Neural Spline Flows (NSF) with a coupling layer \cite{durkan2019neural}. More details about how coupling layers and NSF work can be found in the appendix.

\textbf{Model - SAC. } To compute the best response at each \algo iteration, we use Soft Actor Critic (SAC)~\cite{SACDBLP:journals/corr/abs-1801-01290} (but other PG algorithms would work). SAC is an off-policy algorithm which, as mentioned above, uses the key idea of regularization: instead of considering the objective to simply be the sum of rewards, an entropy term is added to encourage sufficient randomization of the policy and thus address the exploration-exploitation trade-off. To be specific, in our setting, given a population distribution $\mu$, the objective is to maximize:
$
    J_\mu(\pi) = \EE_{(s_t,u_t)} \left[\sum_{t=0}^{+\infty} \gamma^t  r(x_t,v_t,u_t,\mu_t) + \delta \mathcal{H}(\pi(\cdot | s_t))\right],
$
where $\mathcal{H}$ denotes the entropy and $\delta \ge 0$ is a weight. 

To implement the optimization, the SAC algorithm follows the philosophy of actor-critic by training parameterized $Q$-function and policy. To help convergence, the authors of SAC also train a parameterized value function $V$. In practice, the three functions are often approximated by neural networks.

In comparison to other successful methods such as Trust Region Policy Optimization (TRPO)~\cite{DBLP:journals/corr/SchulmanLMJA15} 
or Asynchronous Actor-Critic Agents (A3C), SAC is expected to be more efficient in terms of number of samples required to learn the policy thanks to the use of a replay buffer in the spirit of methods such as Deep Deterministic Policy Gradient (DDPG)~\cite{lillicrap2015continuous}.

\textbf{Metrics. } An issue with studying our flocking model is the absence of a gold standard. Especially, we can not compute the exact exploitability~\cite{perrin2020fictitious} of a policy against a given distribution since we can not compute the exact best response. The exploitability measures how much an agent can gain by replacing its policy $\pi$ with a best response $\pi'$, when the rest of the population plays $\pi$: $\phi(\pi) := \max \limits_{\pi'} J(\mu_0, \pi', \mu^{\pi}) - J(\mu_0, \pi, \mu^{\pi})$. If $\phi(\bar\pi_j) \to 0$ as $j$ increases, FP approaches a Nash equilibrium. 
To cope with these issues, we introduce the following ways to measure progress of the algorithm:
\begin{itemize}
    \item \emph{Performance matrix: } we build the matrix $\mathcal{M}$ of performance of learned policies versus estimated distributions. The entry $\mathcal{M}_{i,j}$ on the $i$-th row and the $j$-th column is the total $\gamma$-discounted sum of rewards: 
$ \mathcal{M}_{i,j} = \mathbb{E}\left[\sum_{t=0}^T \gamma^t r_{t,i} \; | \; s_0 \sim \bar \mu_{i-1}, \; u_t \sim \pi_j(.|s_t)\right],$
where $r_{t,i} = r(s_t, u_t, \bar \mu_{i-1})$, 
     obtained with $\pi_j$ against $\bar\mu_{i-1}$. The diagonal term $\mathcal{M}_{j,j}$ corresponds to the value of the best response computed at iteration $j$.
    
    \item \emph{Approximate exploitability: } We do not have access to the exact best response due to the model-free approach and the continuous spaces. However, we can approximate the first term of $\phi(\bar\pi)$ directly in the \algo algorithm with SAC. The second term, $J(\mu_0, \bar\pi, \mu^{\bar\pi})$, can be approximated by replacing $\bar\pi$ with the average over past policies, {\it i.e.}, the policy sampled uniformly from the set $\{\pi_0, \dots, \pi_j\}$. At step $j$, the approximate exploitability is $e_j = \mathcal{M}_{j,j} - \frac{1}{j-1}\sum_{k=1}^{j-1} \mathcal{M}_{j,k}$. To smoothen the exploitability, we take the best response over the last 5 policies and use a moving average over 10 points. Please note that only relative values are important as it depends on the scale of the reward.
    
\end{itemize}

\paragraph{A 4-Dimensional Example.}
We illustrate in a four dimensional setting (\textit{i.e.} two-dimensional positions and velocities) how the agents learn to adopt similar velocities by controlling their acceleration. 
We focus on the role of $\beta$ in the flocking effect. We consider noise $\epsilon^i_t \sim \mathcal{N}(0,\Delta t)$ and the following reward:
$
    r^i_t= 
     f^{\mathrm{flock},i}_{\beta,t}
     - \|u^i_t\|_2^2 + \|v^i_t\|_{\infty} 
    - \min\{\|x^{i}_{2,t} \pm 50\|\},
$
where $x^{i}_{2,t}$ stands for the second coordinate of the $i$-th agent's position at time $t$. The last term attracts the agents' positions towards one of two lines corresponding to the second coordinate of $x$ being either $-50$ or $+50$. We added a term regarding the norm of the velocity to prevent agents from stopping. Here we take $\|v^i_t\|_{\infty} = \max\{|v^i_{1,t}|, |v^i_{2,t}|\}$. Hence, a possible equilibrium is with two groups of agents, one for each line. When $\beta = 0$, the term $f^{\mathrm{flock},i}_{\beta,t}$ encourages agent $i$ to have the same velocity vector as the rest of the whole population. At equilibrium, the agents in the two groups should thus move in the same direction (to the left or to the right, in order to stay on the two lines of $x$'s). On the other hand, when $\beta>0$ is large enough (\textit{e.g.} $\beta=100$), agent $i$ gives more importance to its neighbors when choosing its control and it tries to have a velocity similar to the agents that are position-wise close to. This allows the emergence of two groups moving in different directions: one group moves towards the left (overall negative velocity) and the other group moves towards the right (overall positive velocity). 

\begin{figure}[tbp]
    
    \begin{subfigure}{0.25\textwidth}
      \centering
      \includegraphics[width=0.9\linewidth]{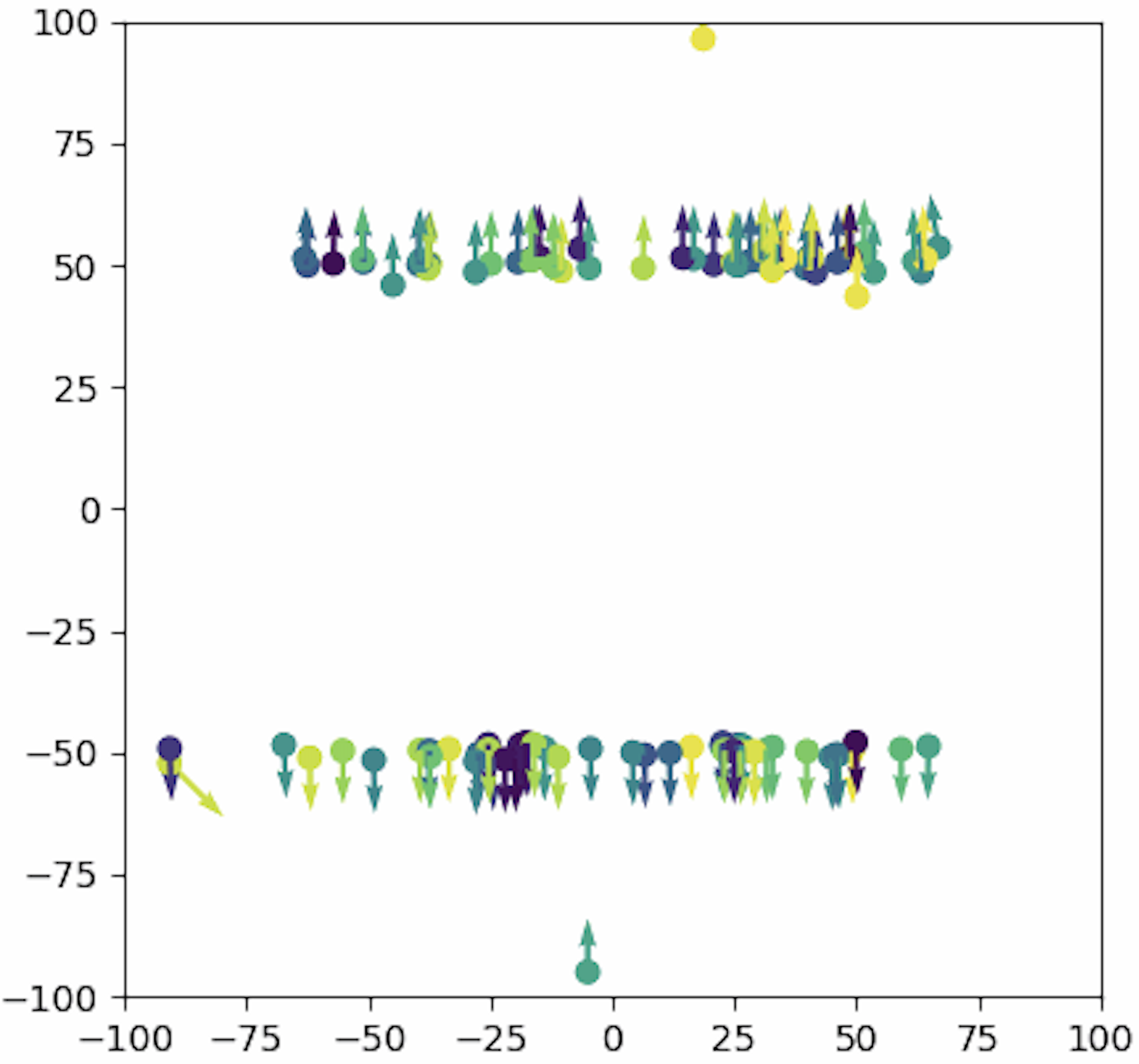}
      \caption{Initial positions and velocities}
    \end{subfigure}%
    \begin{subfigure}{0.25\textwidth}
      \centering
      \includegraphics[width=0.9\linewidth]{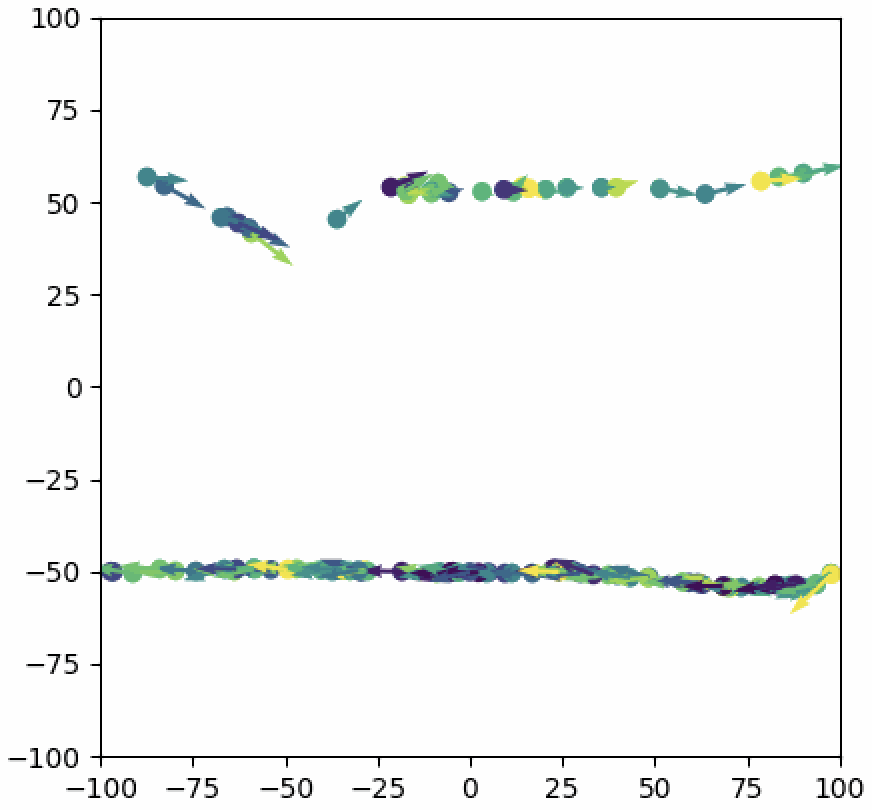}
      \caption{At convergence}
    \end{subfigure}%
    
    \medskip
    
    \begin{subfigure}{0.25\textwidth}
      \centering
      \includegraphics[width=0.9\linewidth]{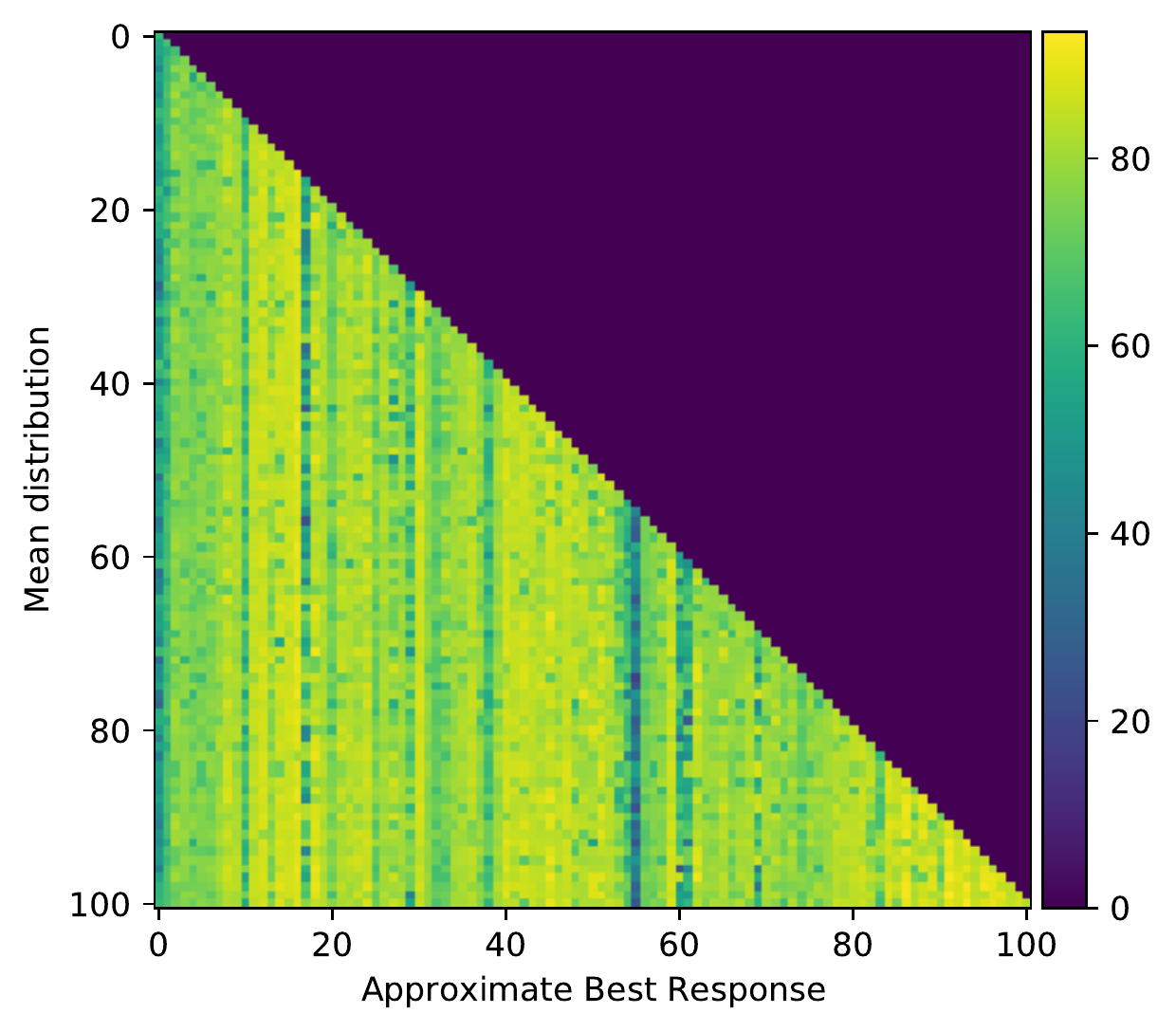}
      \caption{Performance matrix}
    \end{subfigure}%
    \begin{subfigure}{0.25\textwidth}
      \centering
      \includegraphics[width=0.9\linewidth]{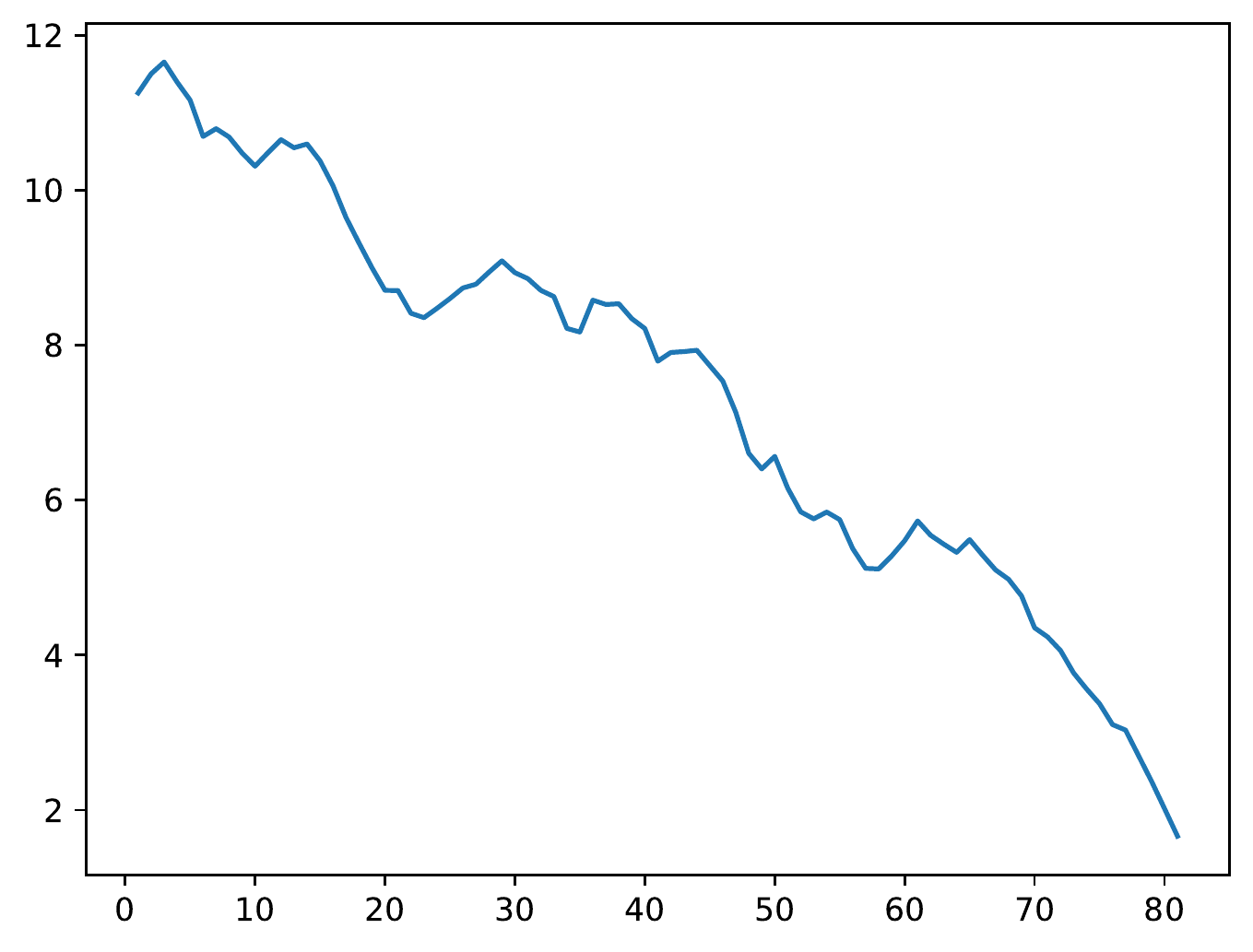}
      \caption{Approximate exploitability}
    \end{subfigure}

    \caption{Multi-group flocking with noise and $\beta=100$.}
    \label{fig:2D_experiment2}
\end{figure}

This is confirmed by Fig.~\ref{fig:2D_experiment2}. In the experiment, we set the initial velocities perpendicular to the desired ones to illustrate the robustness of the algorithm. We observe that the approximate exploitability globally decreases. In the case $\beta=0$, we experimentally verified that there is always a global consensus, \textit{i.e.}, only one line or two lines but moving in the same direction. 

\paragraph{Scaling to 6 Dimensions and non-smooth topology.}

We now present an example with arbitrary obstacles (and thus non-smooth topology) in dimension 6 (position and velocity in dimension 3) which would be very hard to address with classical numerical methods. In this setting, we have multiple columns that the agents are trying to avoid. The reward has the following form:
$
    r^i_t= 
     f^{\mathrm{flock},i}_{\beta,t}
     - \|u^i_t\|_2^2 + \|v^i_t\|_{\infty} 
    - \min\{\|x^{i}_{2,t}\|\} - c*\mathbf{1}_{obs},
$
If an agent hits an obstacle, it gets a negative reward and bounces on it like a snooker ball. After a few iterations, the agents finally find their way through the obstacles. This situation can model birds trying to fly in a city with tall buildings.
In our experiments, we noticed that different random seeds lead to different solutions. This is not surprising as there are a lot of paths that the agents can take to avoid the obstacles and still maximizing the reward function. The exploitability decreases quicker than in the previous experiment. We believe that this is because agents find a way through the obstacles in the first iterations.

\begin{figure}[tbp]

    \begin{subfigure}{0.25\textwidth}
      \centering
      \includegraphics[width=1.0\linewidth]{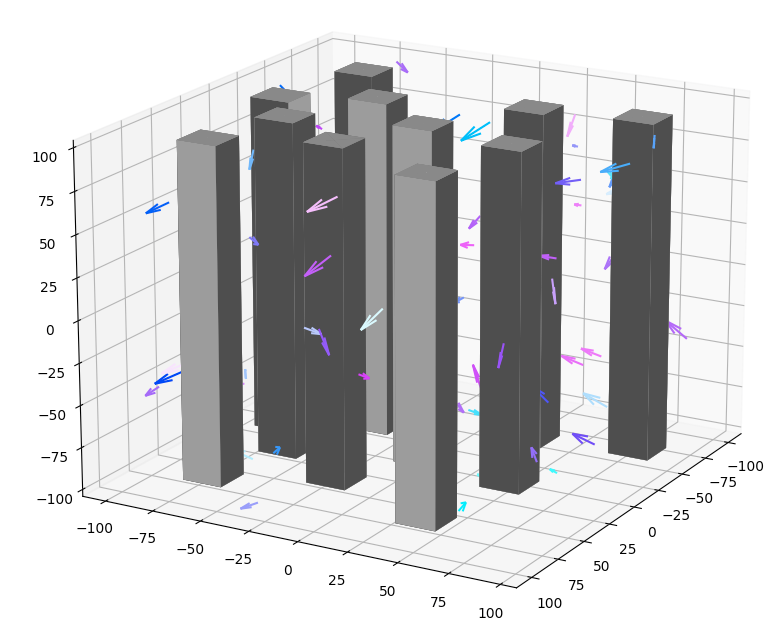}
      \caption{Initial positions and velocities}
    \end{subfigure}%
    \begin{subfigure}{0.25\textwidth}
      \centering
      \includegraphics[width=1.0\linewidth]{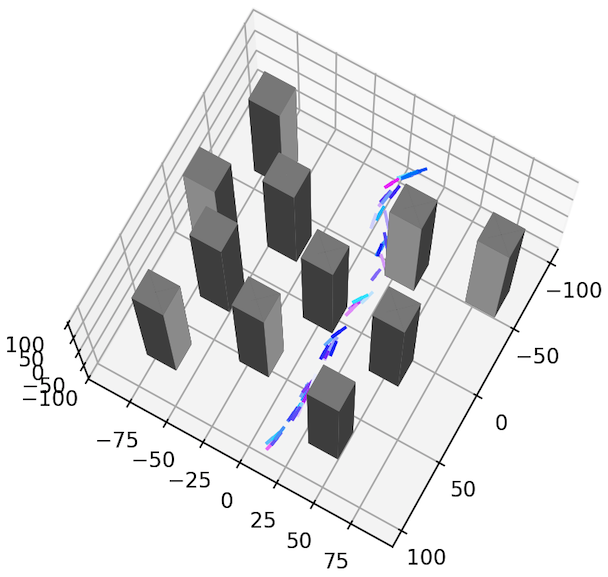}
      \caption{At convergence}
    \end{subfigure}%
    
    \medskip
    
    \begin{subfigure}{0.25\textwidth}
      \centering
      \includegraphics[width=0.9\linewidth]{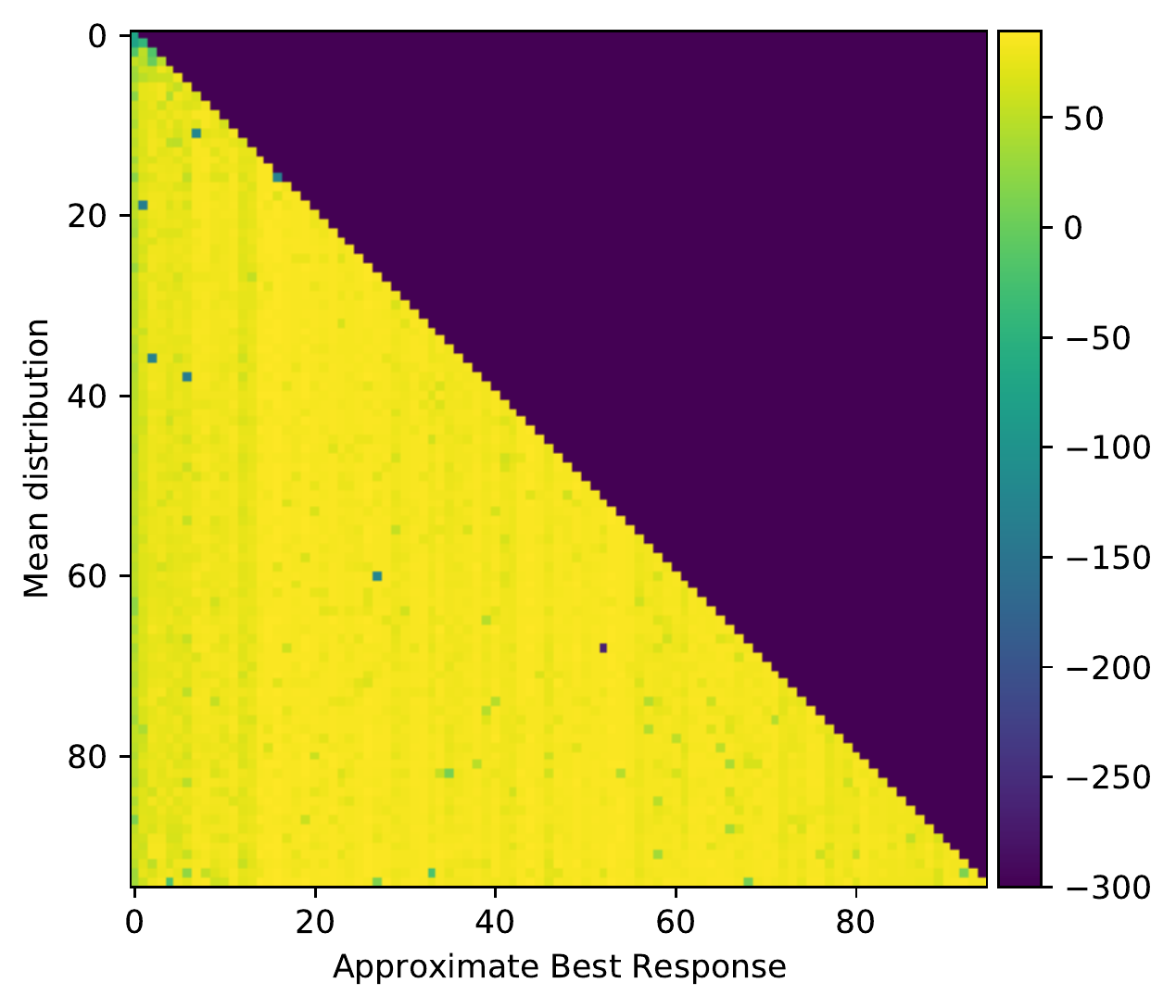}
      \caption{Performance matrix}
    \end{subfigure}%
    \begin{subfigure}{0.25\textwidth}
      \centering
      \includegraphics[width=0.9\linewidth]{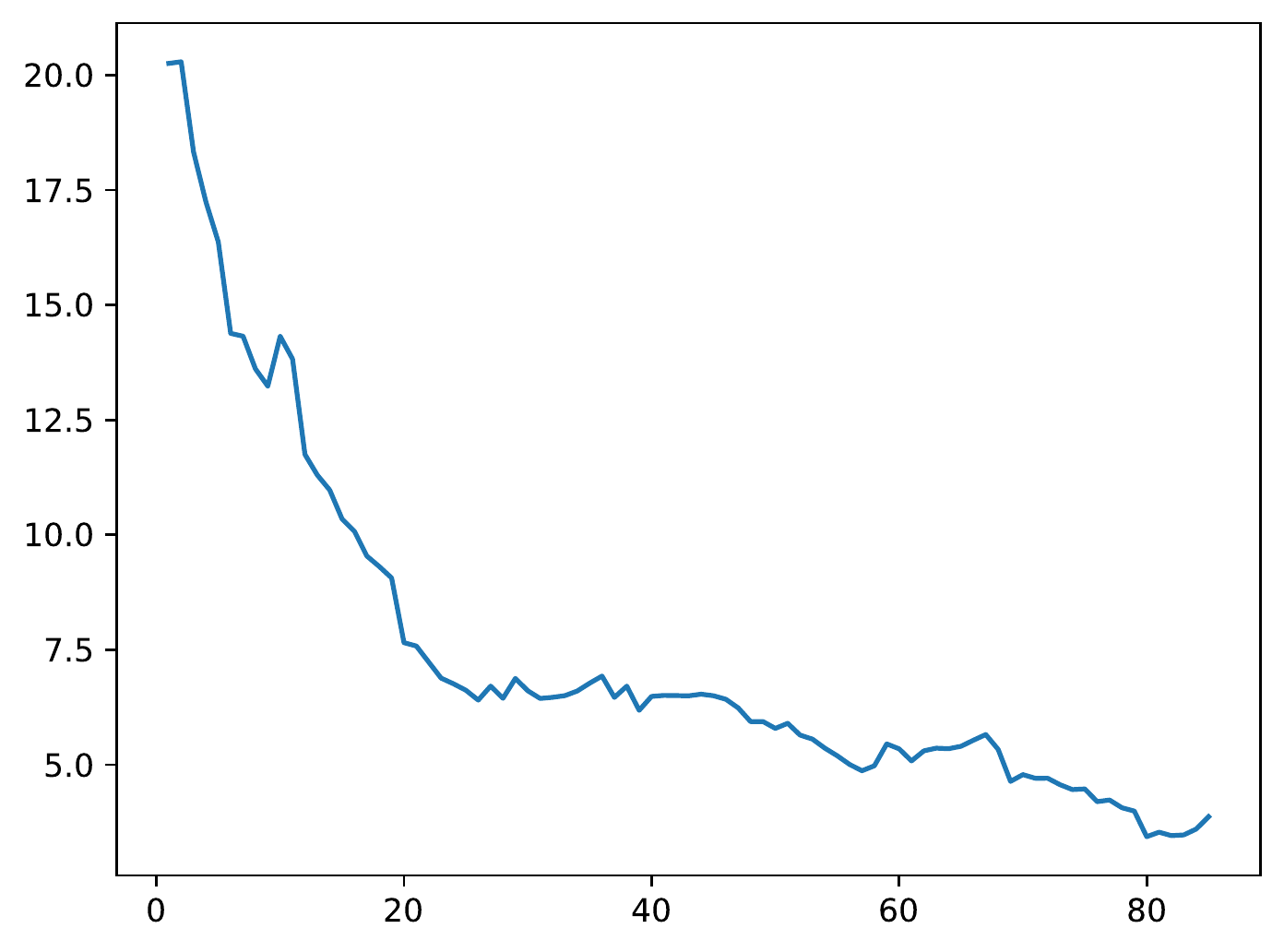}
      \caption{Approximate exploitability}
    \end{subfigure}

    \caption{Flocking with noise and many obstacles.}
    \label{fig:3D_building}
\end{figure}

\section{Related Work}

\textbf{Numerical methods for flocking models. } Most work using flocking models focus on the dynamical aspect without optimization. To the best of our knowledge, the only existing numerical approach to tackle a MFG with flocking effects is in~\cite[Section 4.7.3]{CarmonaDelarue_book_I}, but it is restricted to a very special and simpler type of rewards. 

\noindent
\textbf{Learning in MFGs. }
MFGs have attracted a surge of interest in the RL community as a possible way to remediate the scalability issues encountered in MARL when the number of agents is large~\cite{perolat2018actor}. \cite{guo2019learning} combined a fixed-point method with $Q$-learning, but the convergence is ensured only under very restrictive Lipschitz conditions and the method can be applied efficiently only to finite-state models. \cite{subramanianpolicy} solve MFG using a gradient-type approach. 
The idea of using FP in MFGs has been introduced in~\cite{MR3608094}, assuming the agent can compute perfectly the best response.  \cite{elie2020convergence,perrin2020fictitious} combined FP with RL methods. However, the numerical techniques used therein do not scale to higher dimensions.

\section{Conclusion}

In this work we introduced \algo, a new numerical approach which allows solving MFGs with flocking effects where the agents reach a consensus in a decentralized fashion. \algo combines Fictitious Play with deep neural networks and reinforcement learning techniques (normalizing flows and soft actor-critic). We illustrated the method on challenging examples, for which no solution was previously known. In the absence of existing benchmark, we demonstrated the success of the method using a new kind of approximate exploitability. Thanks to the efficient representation of the distribution and to the model-free computation of a best response, the techniques developed here could be used to solve other acceleration controlled MFGs~\cite{achdou2020deterministic} 
or, more generally, other high-dimensional MFGs. Last, the flexibility of RL, which does not require a perfect knowledge of the model, allow us to tackle MFGs with complex topologies (such as boundary conditions or obstacles), which is a difficult problem for traditional methods based on partial differential equations.

\clearpage

\bibliographystyle{named}
\bibliography{biblio_ijcai}

\appendix

\clearpage

\section{More numerical tests}

\subsection{A Simple Example in Four dimensions}

We illustrate in a simple four dimensional setting (\textit{i.e.} two-dimensional positions and velocities hence the total dimension is $4$) how the agents learn to adopt similar velocities by controlling their acceleration.

Here, we take $\epsilon^i_t \equiv 0$ (no noise), and we define the reward as:
\begin{equation}
    r^i_t = 
    f^{\mathrm{flock},i}_{\beta=0,t}
    - \|u^i_t\|_2^2 + \|v^i_t\|_2^2.
    \label{eq_reward_test_case1}
\end{equation}
We set $\beta=0$ to have a intuitive example. The first term encourages the agent to adapt its velocity to the crowd's one, giving equal importance to all the agents irrespective of their distance.  The second term penalizes a strong acceleration and we added the last term (not present in the original flocking model) to reduce the number of Nash equilibria and prevent the agent to converge to a degenerate solution which consists in putting the whole crowd at a common position with a null velocity.

As the velocity is bounded by 1 in our experiments, there are at least four obvious Nash equilibria in terms of velocity that remain from reward \eqref{eq_reward_test_case1}: $\hat v^i_t \equiv (-1,-1)$, $(-1,1)$, $(1,-1)$ and $(1,1)$, while the position is not important anymore and any distribution for the positions is valid. Experimentally, we observe that if we start from a normal initial distribution $v^i_0 \sim \mathcal{N}(0,1)$, then the equilibrium found by \algo is randomly one of the four previous velocities. However, if we set the initial velocities with a positive or negative bias, then we observe experimentally that the corresponding equilibrium is reached. Thus, as we could expect, the Nash equilibria found by the algorithm depends on the initial distribution.

\begin{figure}[htbp]
    \centering
    \begin{subfigure}{0.25\textwidth}
      \centering
      \includegraphics[width=0.9\linewidth]{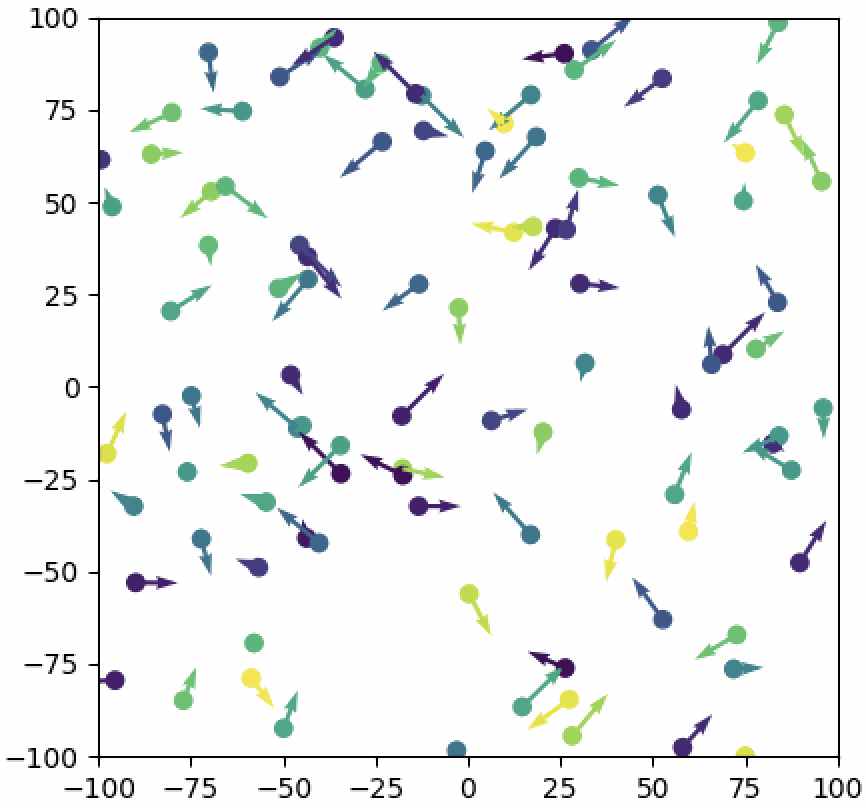}
      \caption{Initial positions and velocities}
    \end{subfigure}%
    \begin{subfigure}{0.25\textwidth}
      \centering
      \includegraphics[width=0.9\linewidth]{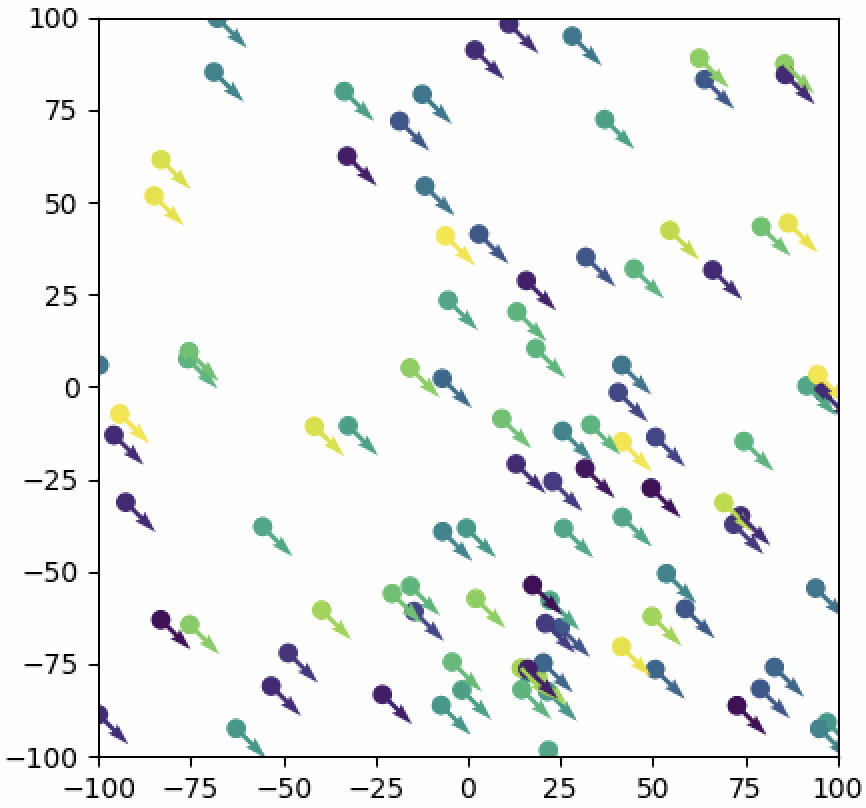}
      \caption{At convergence}
    \end{subfigure}%
    
    \medskip
    
    \begin{subfigure}{0.25\textwidth}
      \centering
      \includegraphics[width=0.9\linewidth]{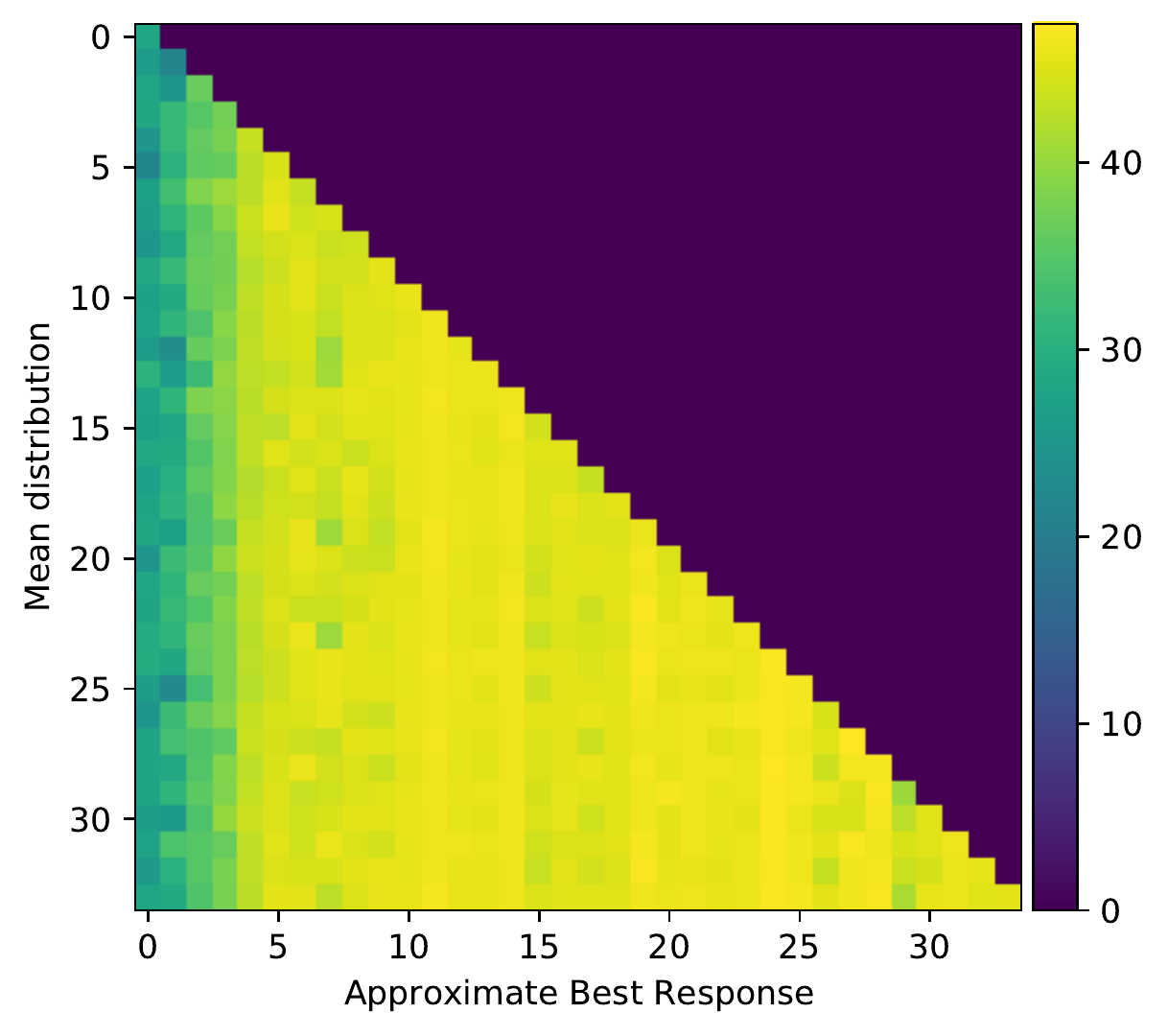}
      \caption{Performance matrix}
    \end{subfigure}%
    \begin{subfigure}{0.25\textwidth}
      \centering
      \includegraphics[width=0.9\linewidth]{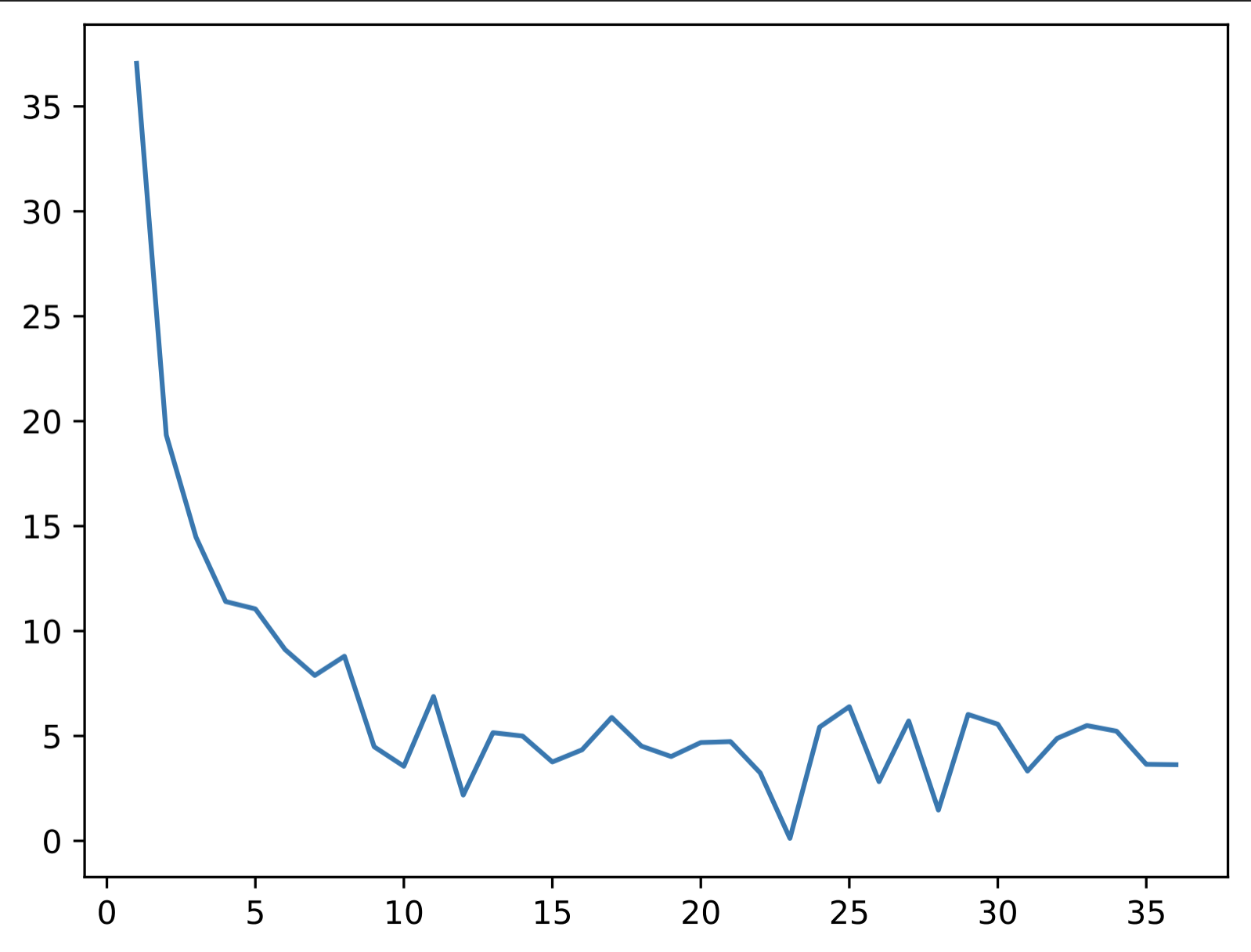}
      \caption{Approximate exploitability}
    \end{subfigure}

    \caption{Flocking with an intuitive example}
    \label{fig:2D_experiment_test1}
\end{figure}

In Fig.~\ref{fig:2D_experiment_test1}, we can see that the agents have adopted the same velocities and the performance matrix indicates that the policy learned at each step of \algo tends to perform better than previous policies. The exploitablity decreases quickly because \algo algorithm learns during the first iterations an approximation of a Nash equilibrium.

\subsection{A Simple Example in Six Dimensions}

\begin{figure}[htbp]
    
    \begin{subfigure}{0.25\textwidth}
      \centering
      \includegraphics[width=1.0\linewidth]{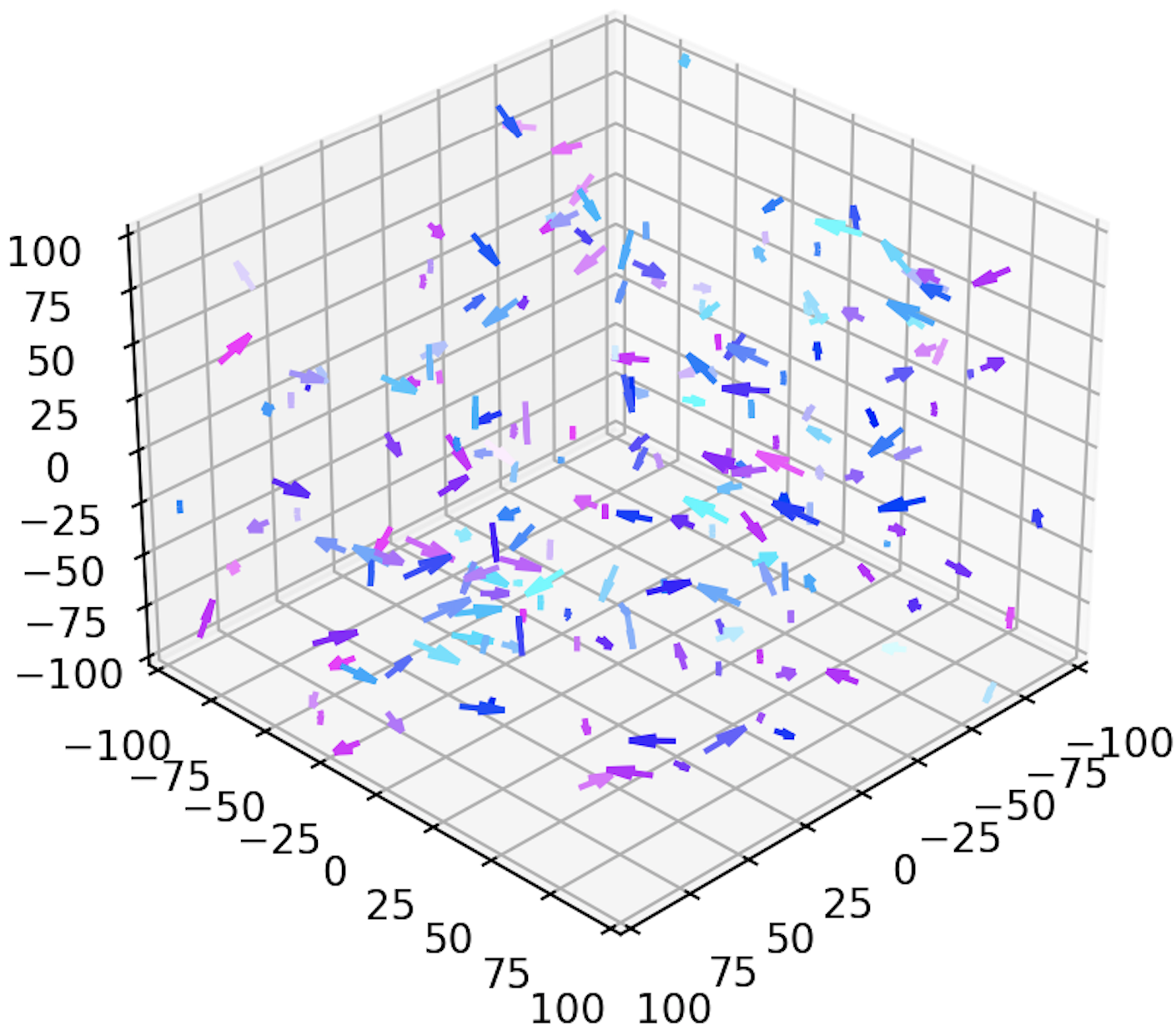}
      \caption{Initial positions and velocities}
      \label{fig:init_3D_experiment_test_case_1}
    \end{subfigure}%
    \begin{subfigure}{0.25\textwidth}
      \centering
      \includegraphics[width=1.0\linewidth]{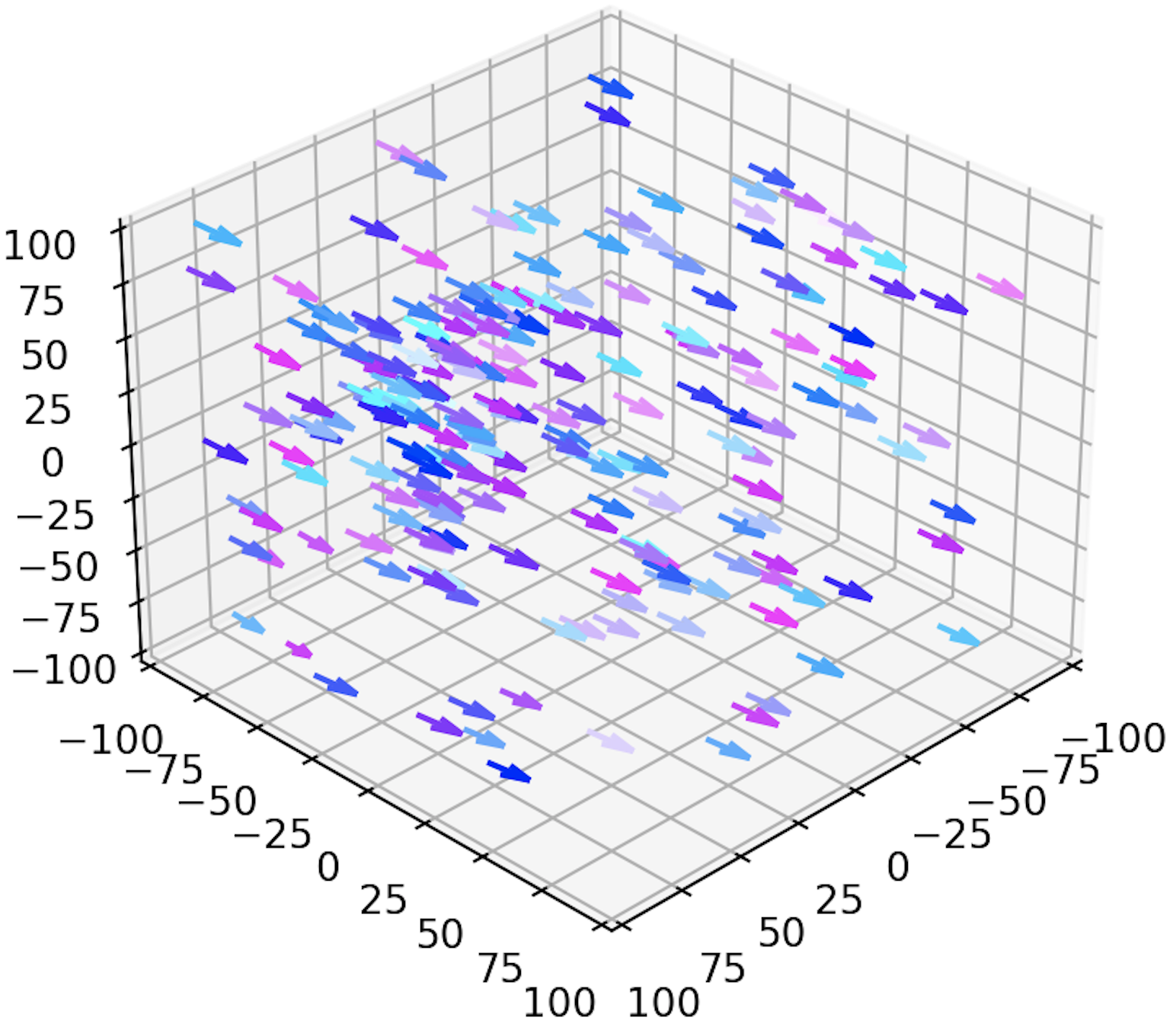}
      \caption{At convergence}
      \label{fig:converge_3D_experiment_test_case_1}
    \end{subfigure}%
    
    \medskip
    
    \begin{subfigure}{0.25\textwidth}
      \centering
      \includegraphics[width=0.9\linewidth]{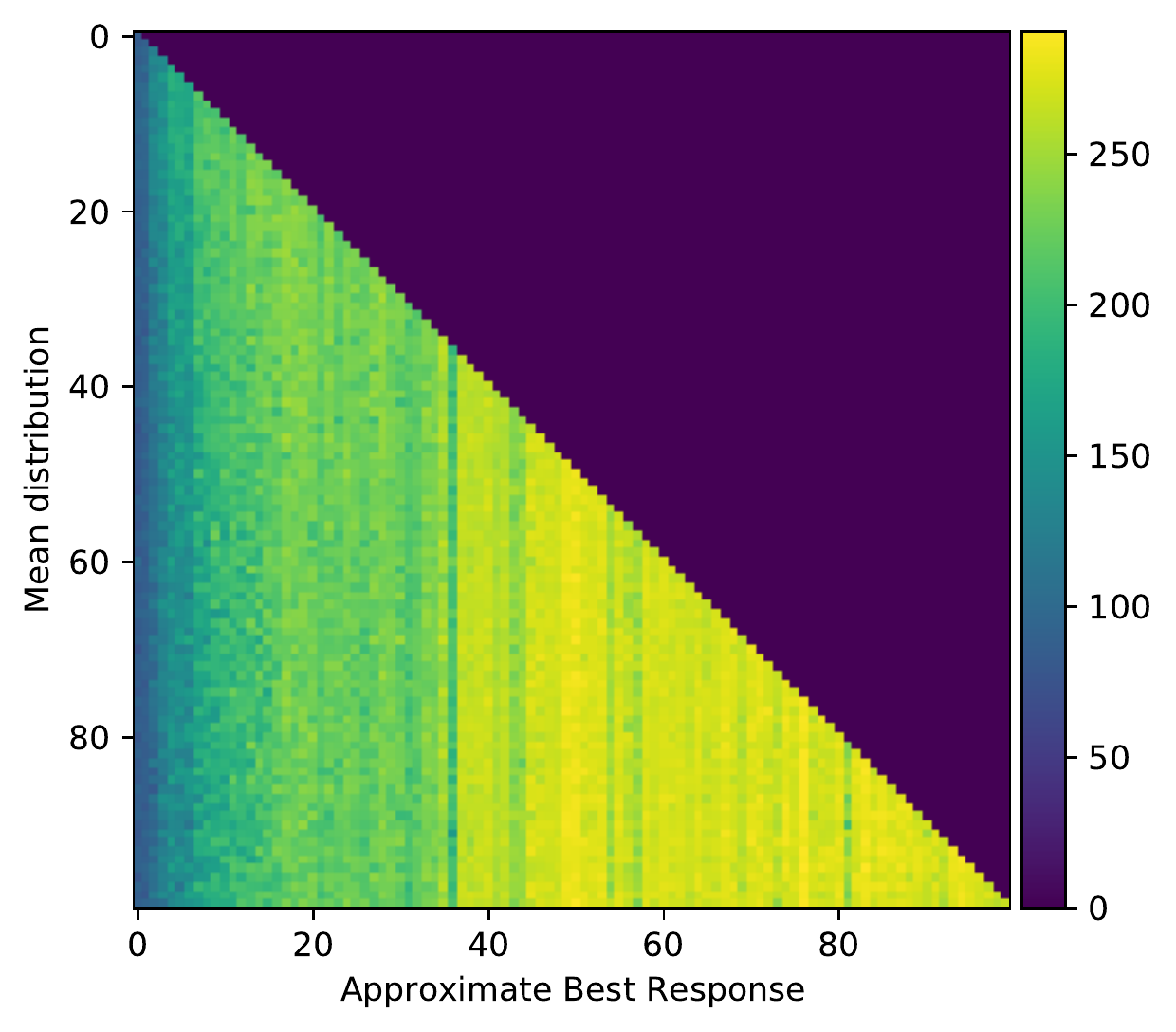}
      \caption{Performance matrix}
      \label{fig:pm_3D_experiment_test_case_1}
    \end{subfigure}%
    \begin{subfigure}{0.25\textwidth}
      \centering
      \includegraphics[width=0.9\linewidth]{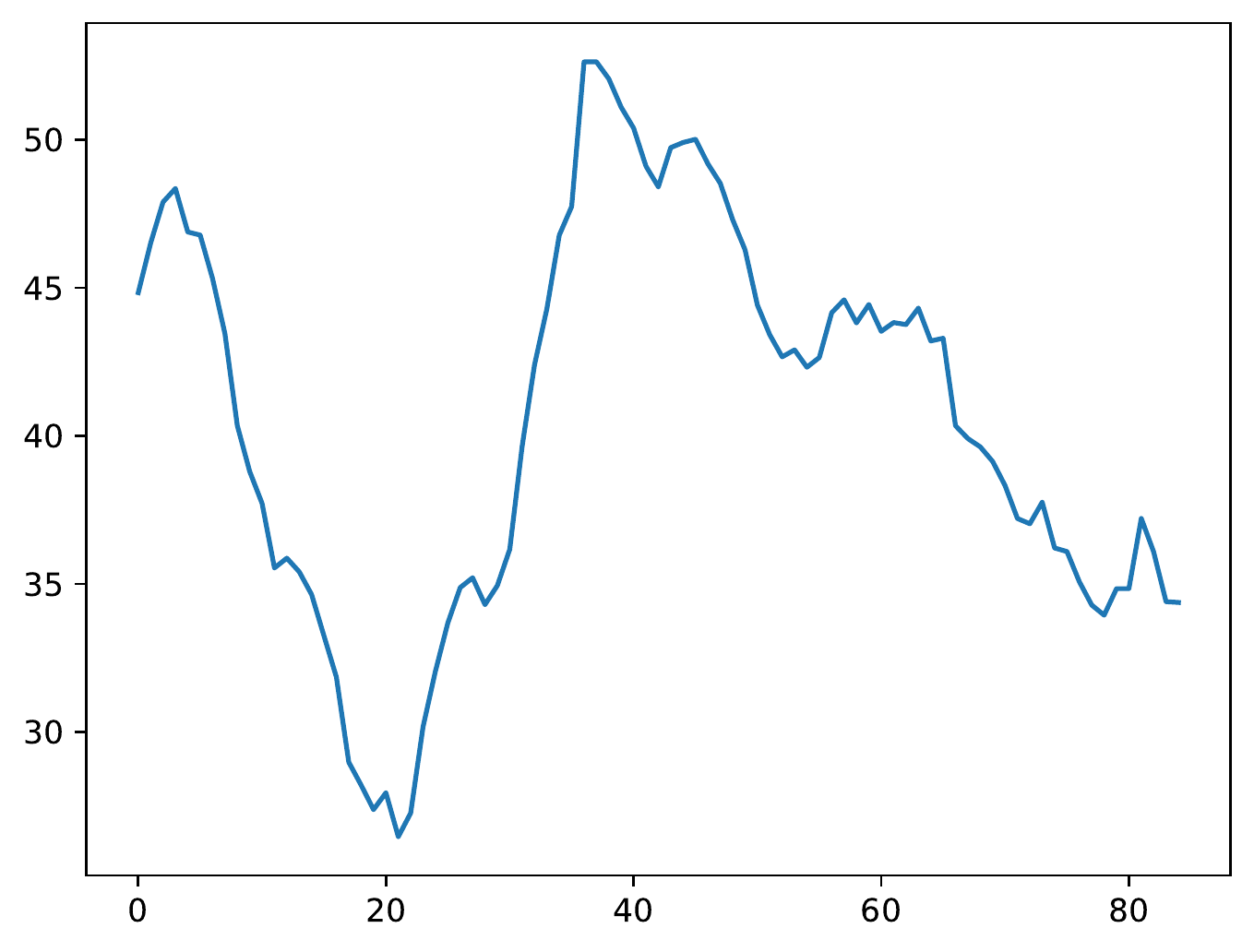}
      \caption{Approximate exploitability}
      \label{fig:ae_3D_experiment_test_case_1}
    \end{subfigure}

    \caption{Flocking for a simple example in dimension 6.}
    \label{fig:3D_experiment_test_case_1}
\end{figure}

We consider the simple example defined with reward from Eq.~\eqref{eq_reward_test_case1} but now we add an additional dimension for the position and the velocity, making the problem six-dimensional. As before, the agents are still encouraged to maximize their velocities. We set $\beta=0$ and we do not put any noise on the velocity dynamics. We can notice that experimentally a consensus is reached by the agents as they learn to adopt the same velocities  (Fig.~\ref{fig:converge_3D_experiment_test_case_1}), even if the agents start from a random distribution in terms of positions and velocities (Fig.~\ref{fig:init_3D_experiment_test_case_1}). The performance matrix (Fig.~\ref{fig:pm_3D_experiment_test_case_1}) highlights that the best response improves until iteration 40, which explains why there is a bump in performance in the exploitability before the 40th iteration.

\subsection{Examples with an obstacle}

\begin{figure}[htbp]
    
    \begin{subfigure}{0.25\textwidth}
      \centering
      \includegraphics[width=0.8\linewidth]{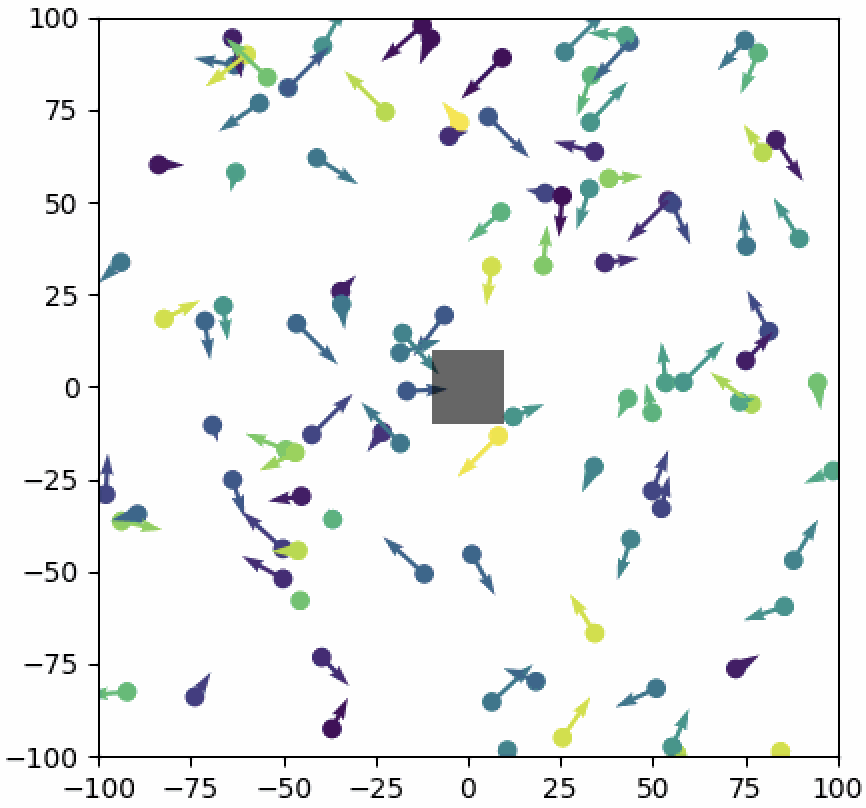}
      \caption{Initial positions and velocities}
    \end{subfigure}%
    \begin{subfigure}{0.25\textwidth}
      \centering
      \includegraphics[width=0.8\linewidth]{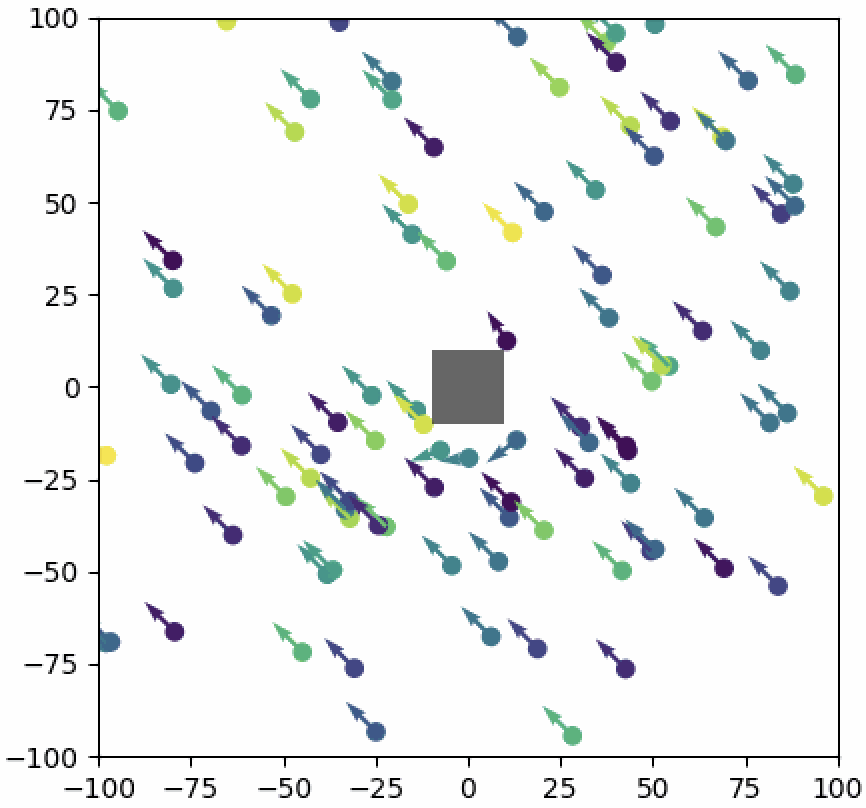}
      \caption{At convergence}
      \label{fig:converge_4d_one_obstacle}
    \end{subfigure}%
    
    \medskip
    
    \begin{subfigure}{0.25\textwidth}
      \centering
      \includegraphics[width=0.9\linewidth]{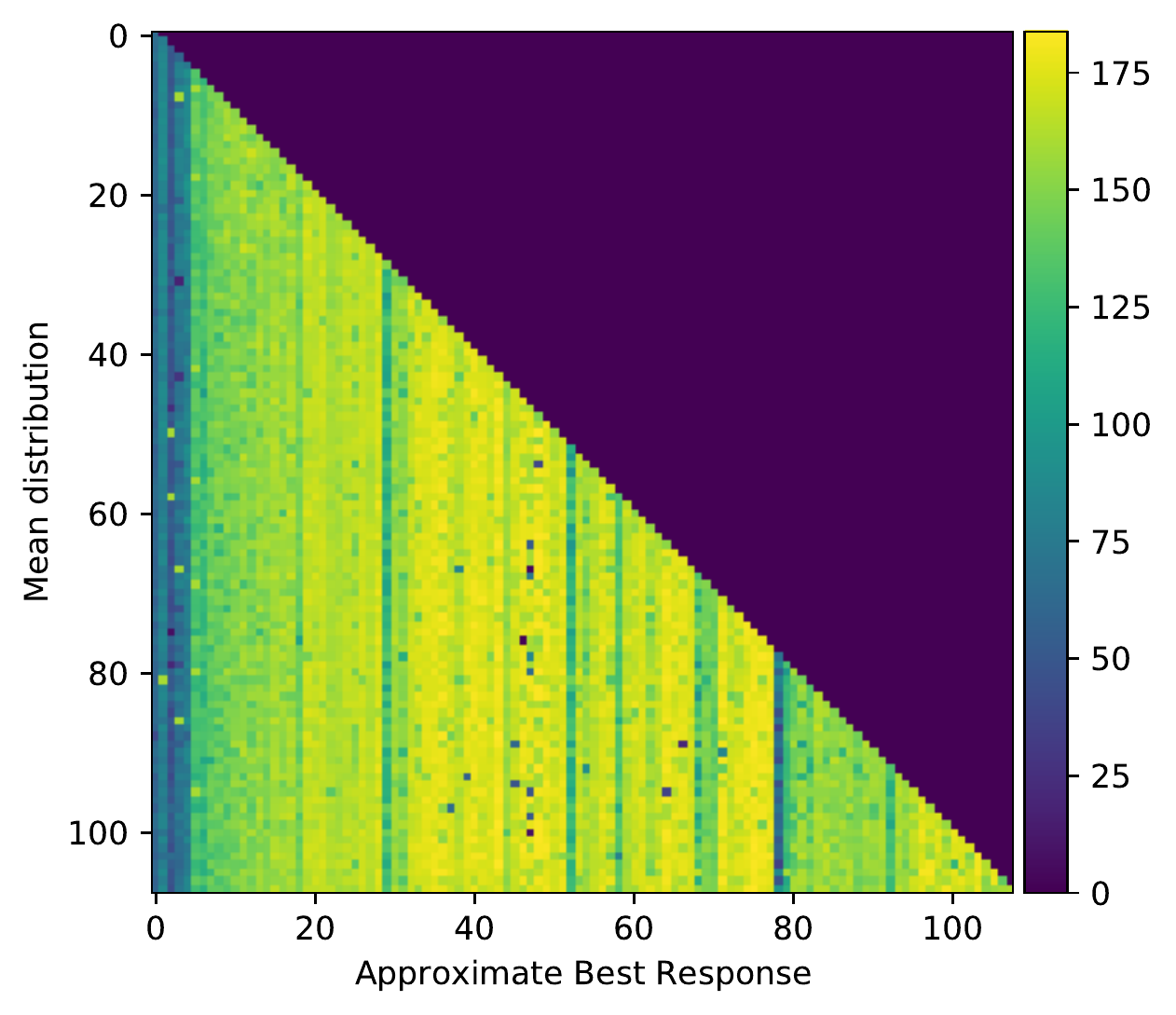}
      \caption{Performance matrix}
    \end{subfigure}%
    \begin{subfigure}{0.25\textwidth}
      \centering
      \includegraphics[width=0.9\linewidth]{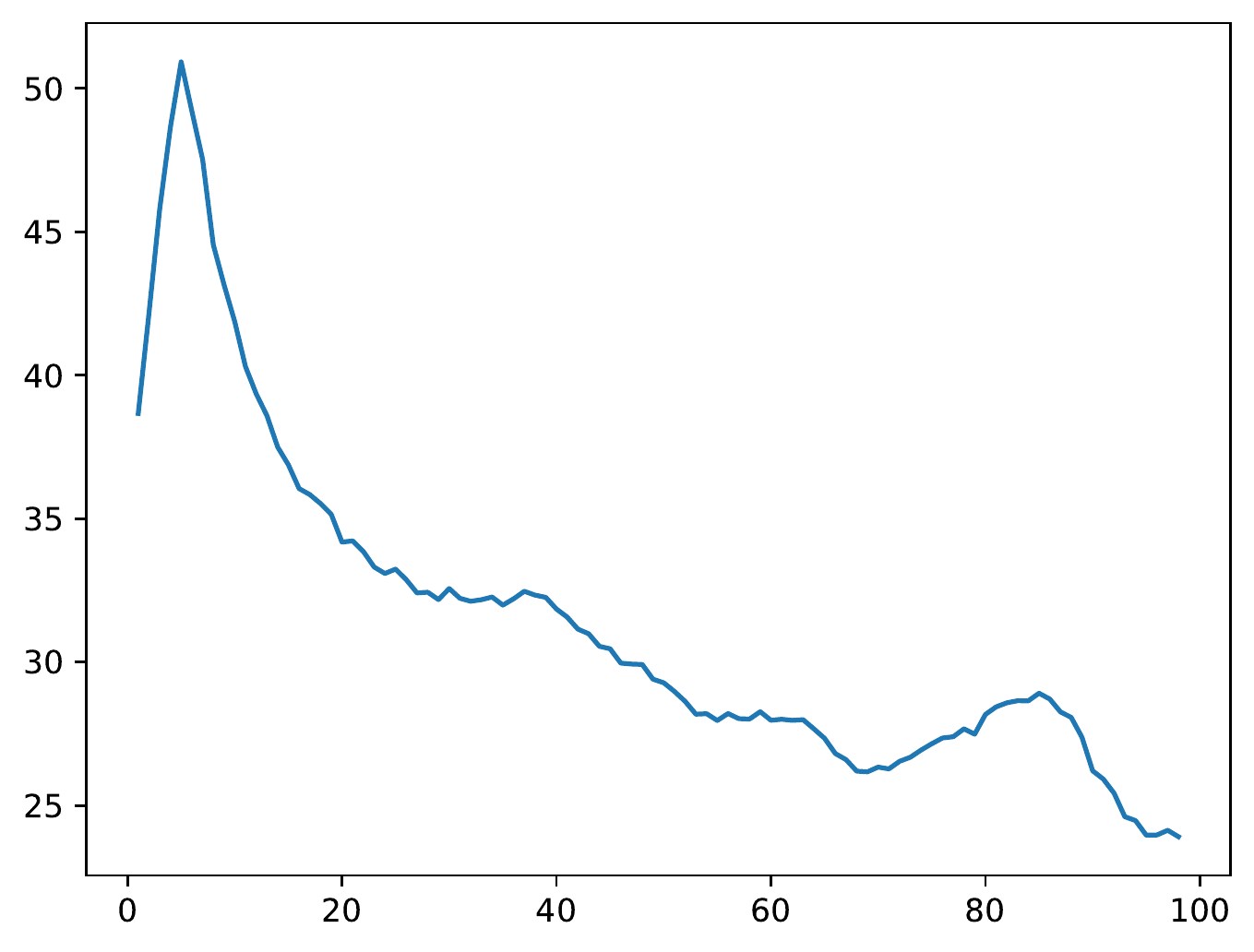}
      \caption{Approximate exploitability}
    \end{subfigure}

    \caption{Flocking in 4D with one obstacle.}
    \label{fig:4d_one_obstacle}
\end{figure}

We present an example with a single obstacle and noise in the dynamics, both in dimension 4 and 6. In dimension 4, the obstacle is a square located at the middle of the environment, whereas in dimension 6 it is a cube. In dimension 4, we can see in Fig.~\ref{fig:4d_one_obstacle} that the agents that are initially spawned in the environment with random positions and velocities manage to learn to adopt the same velocities while avoiding the obstacle (Fig.~\ref{fig:converge_4d_one_obstacle}). The same behavior is observed in dimension 6. We also notice that the exploitability is slower to decrease in dimension 6, similarly to what we observed with the simple example without the obstacle.

\begin{figure}[htbp]
    
    \begin{subfigure}{0.25\textwidth}
      \centering
      \includegraphics[width=1.0\linewidth]{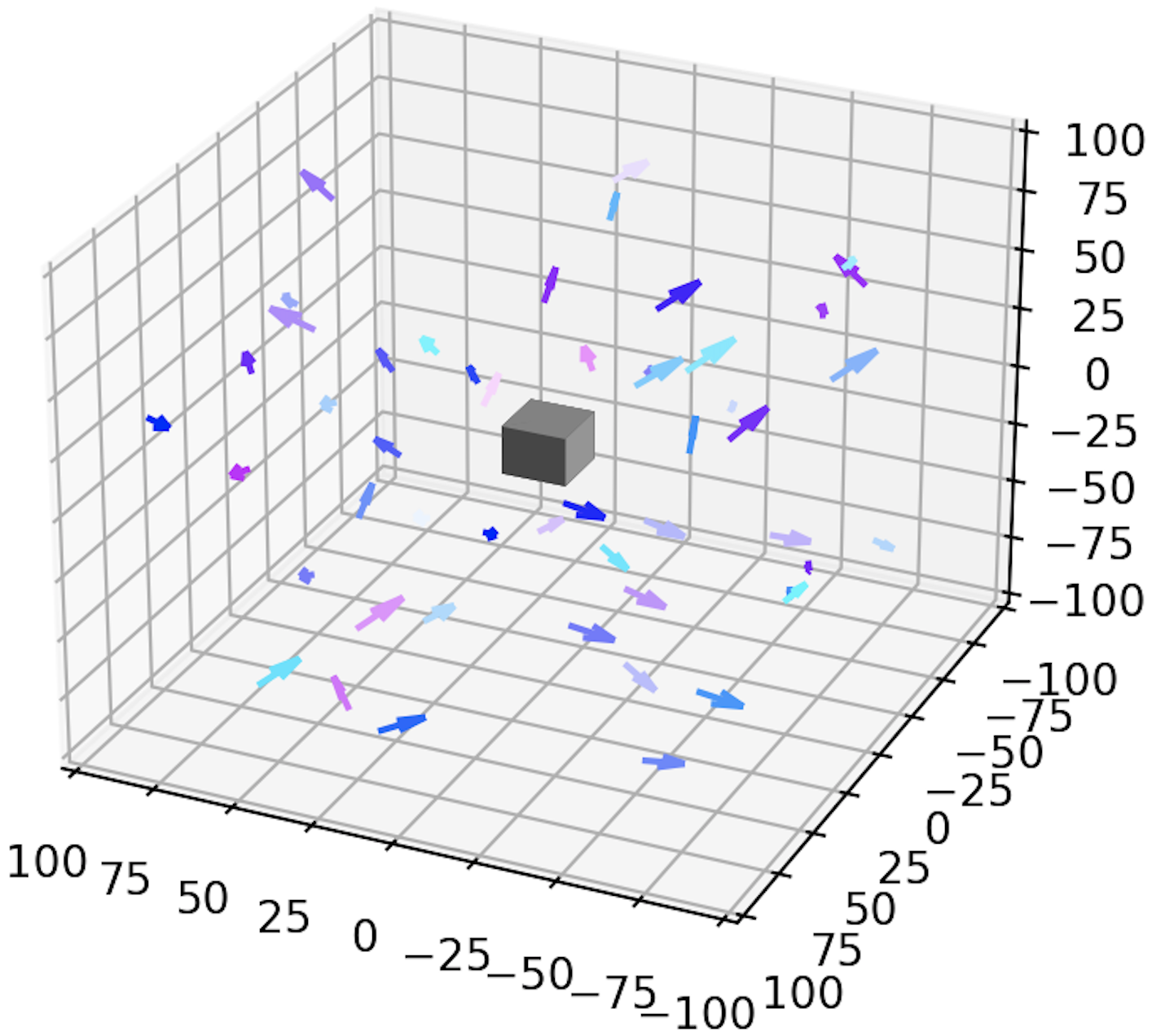}
      \caption{Initial positions and velocities}
    \end{subfigure}%
    \begin{subfigure}{0.25\textwidth}
      \centering
      \includegraphics[width=1.0\linewidth]{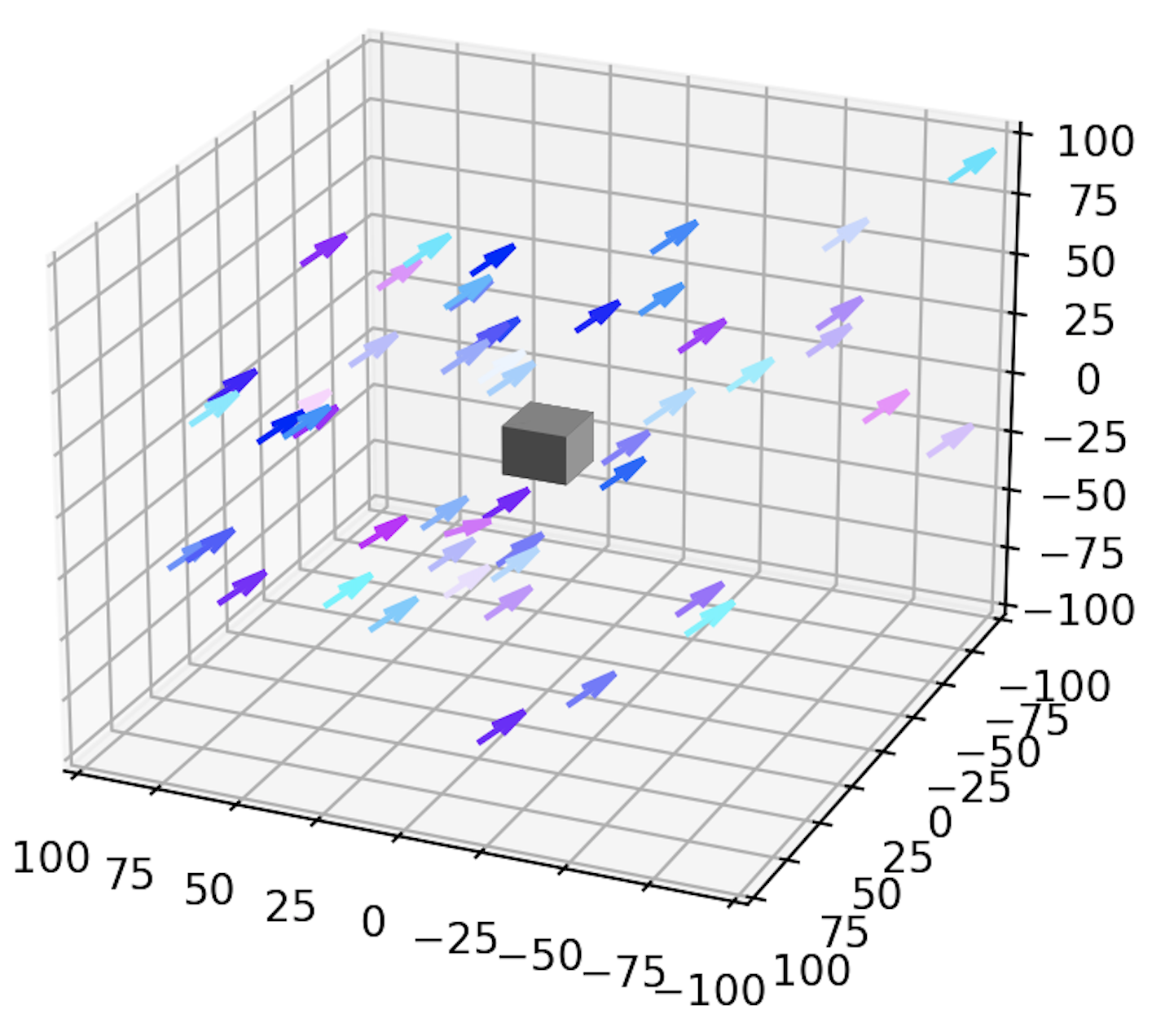}
      \caption{At convergence}
    \end{subfigure}%
    
    \medskip
    
    \begin{subfigure}{0.25\textwidth}
      \centering
      \includegraphics[width=0.9\linewidth]{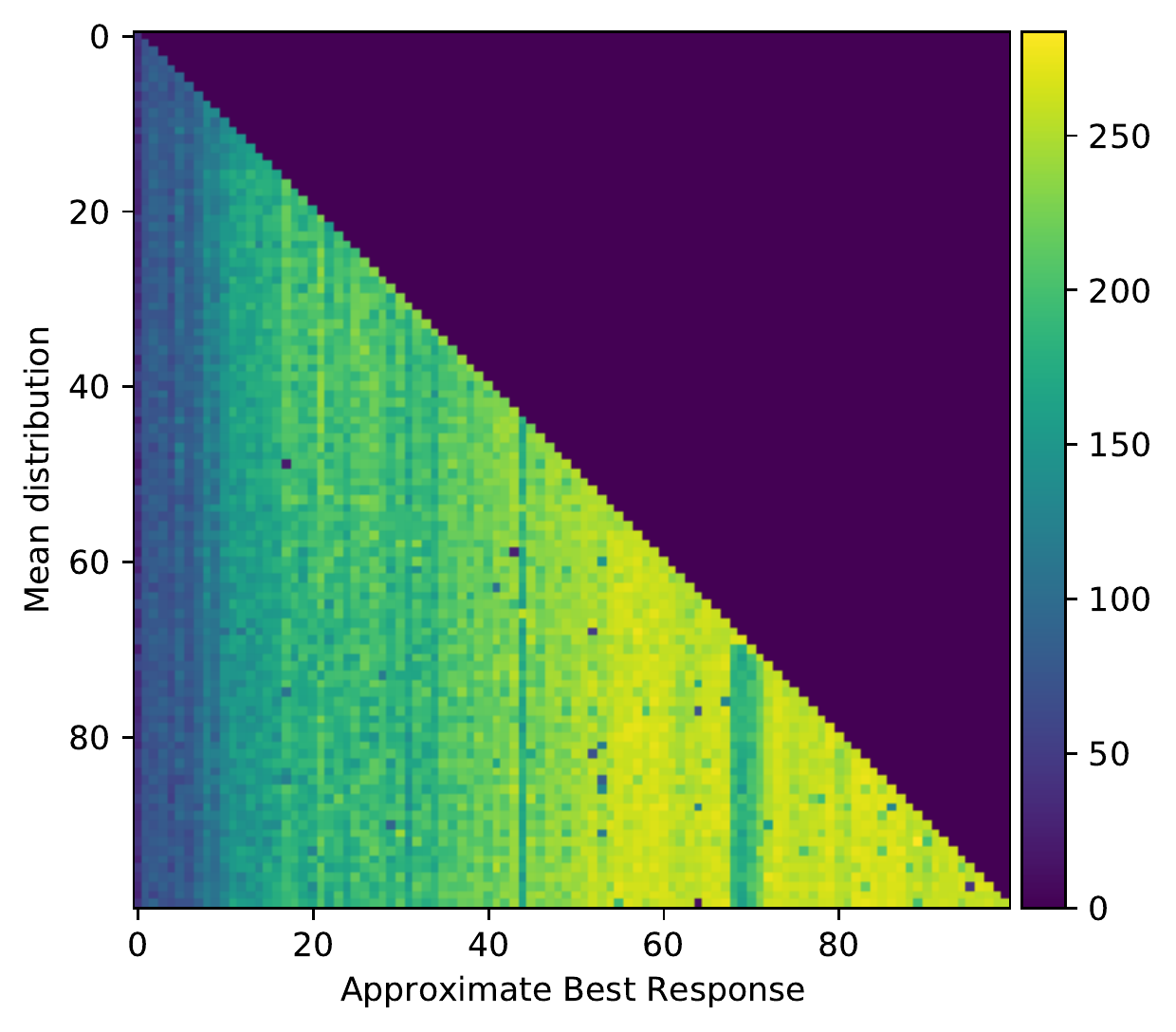}
      \caption{Performance matrix}
    \end{subfigure}%
    \begin{subfigure}{0.25\textwidth}
      \centering
      \includegraphics[width=0.9\linewidth]{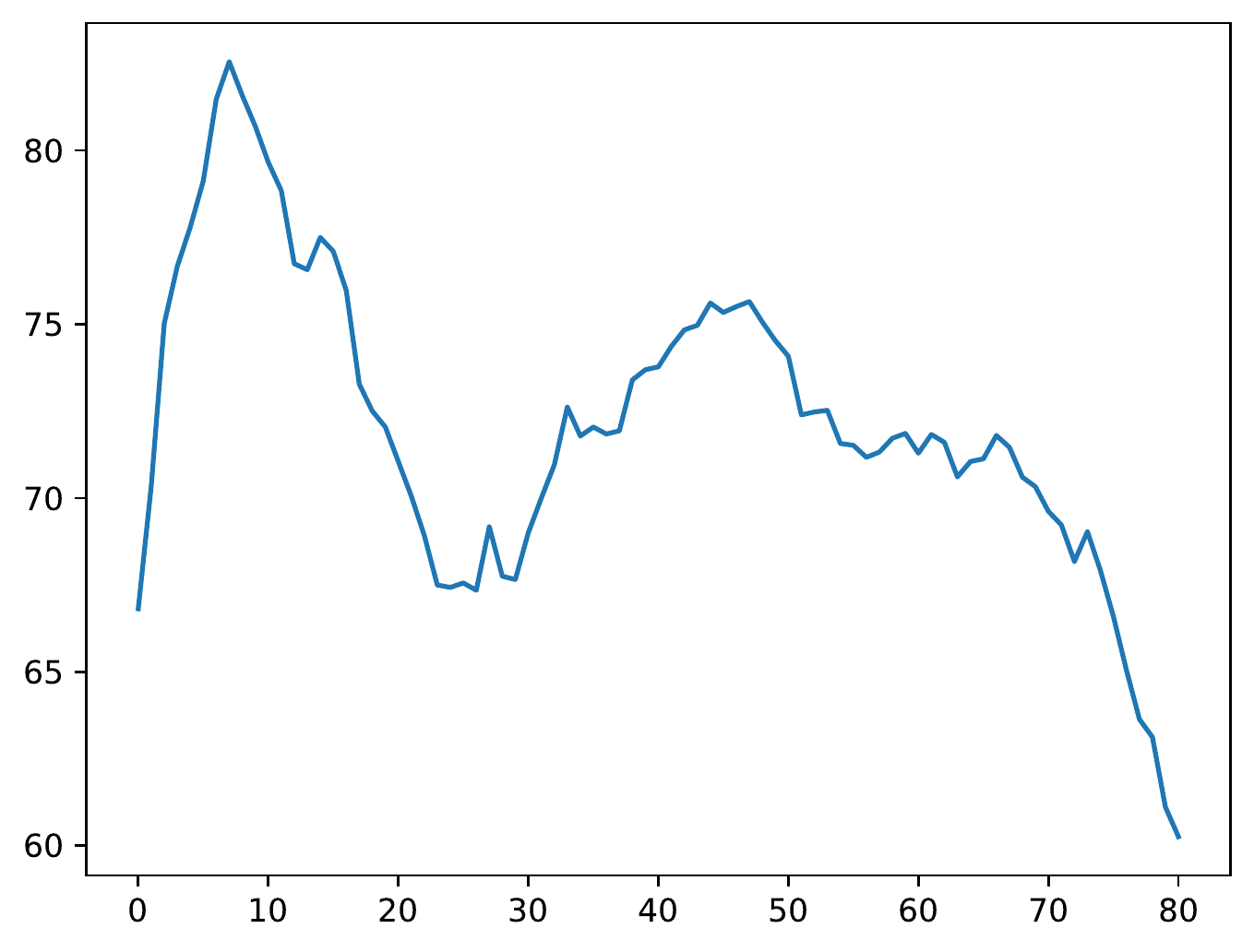}
      \caption{Approximate exploitability}
    \end{subfigure}

    \caption{Flocking in 6D with one obstacle.}
    \label{fig:6d_one_obstacle}
\end{figure}

\subsection{Many obstacles in 4D}

Finally, we present an example in 4D with many obstacles, in the same fashion than the 6-dimensional example with the columns located in the main part of the article. 

\begin{figure}[htbp]
    
    \begin{subfigure}{0.25\textwidth}
      \centering
      \includegraphics[width=0.8\linewidth]{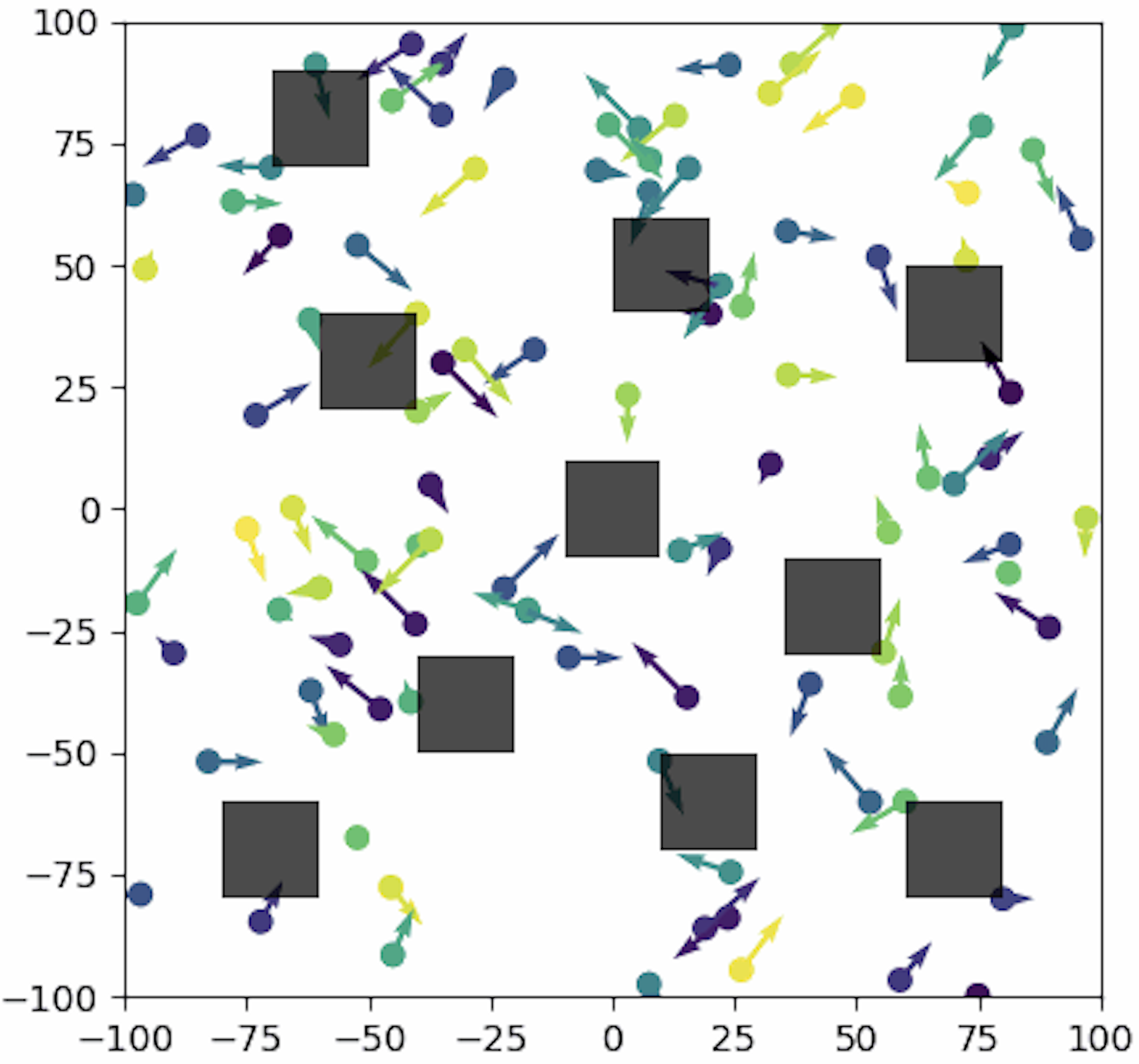}
      \caption{Initial positions and velocities}
    \end{subfigure}%
    \begin{subfigure}{0.25\textwidth}
      \centering
      \includegraphics[width=0.8\linewidth]{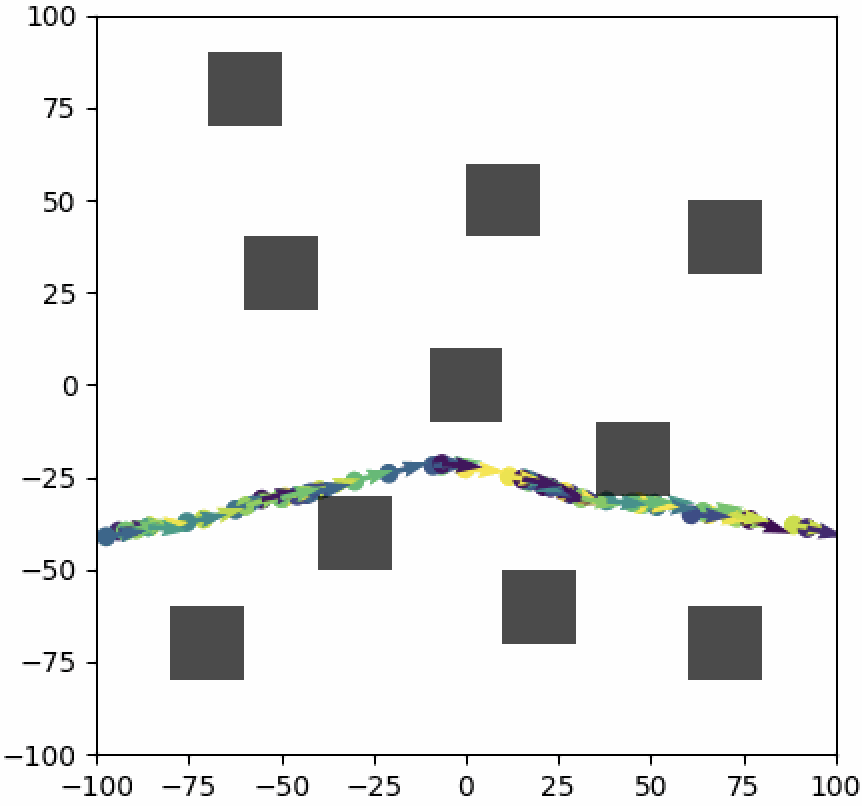}
      \caption{At convergence}
      \label{fig:converge_4d_many_obstacle}
    \end{subfigure}%
    
    \medskip
    
    \begin{subfigure}{0.25\textwidth}
      \centering
      \includegraphics[width=0.9\linewidth]{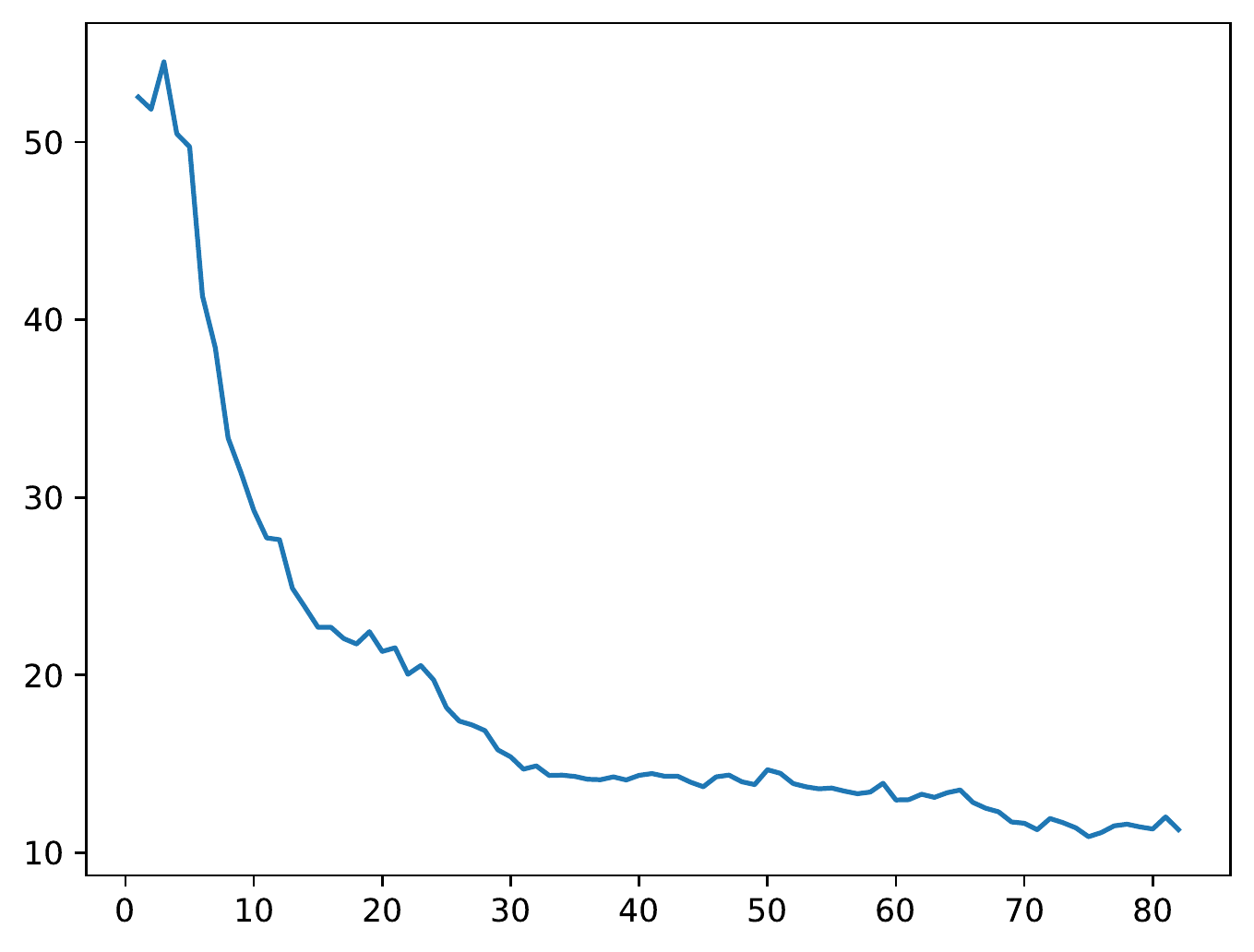}
      \caption{Performance matrix}
    \end{subfigure}%
    \begin{subfigure}{0.25\textwidth}
      \centering
      \includegraphics[width=0.9\linewidth]{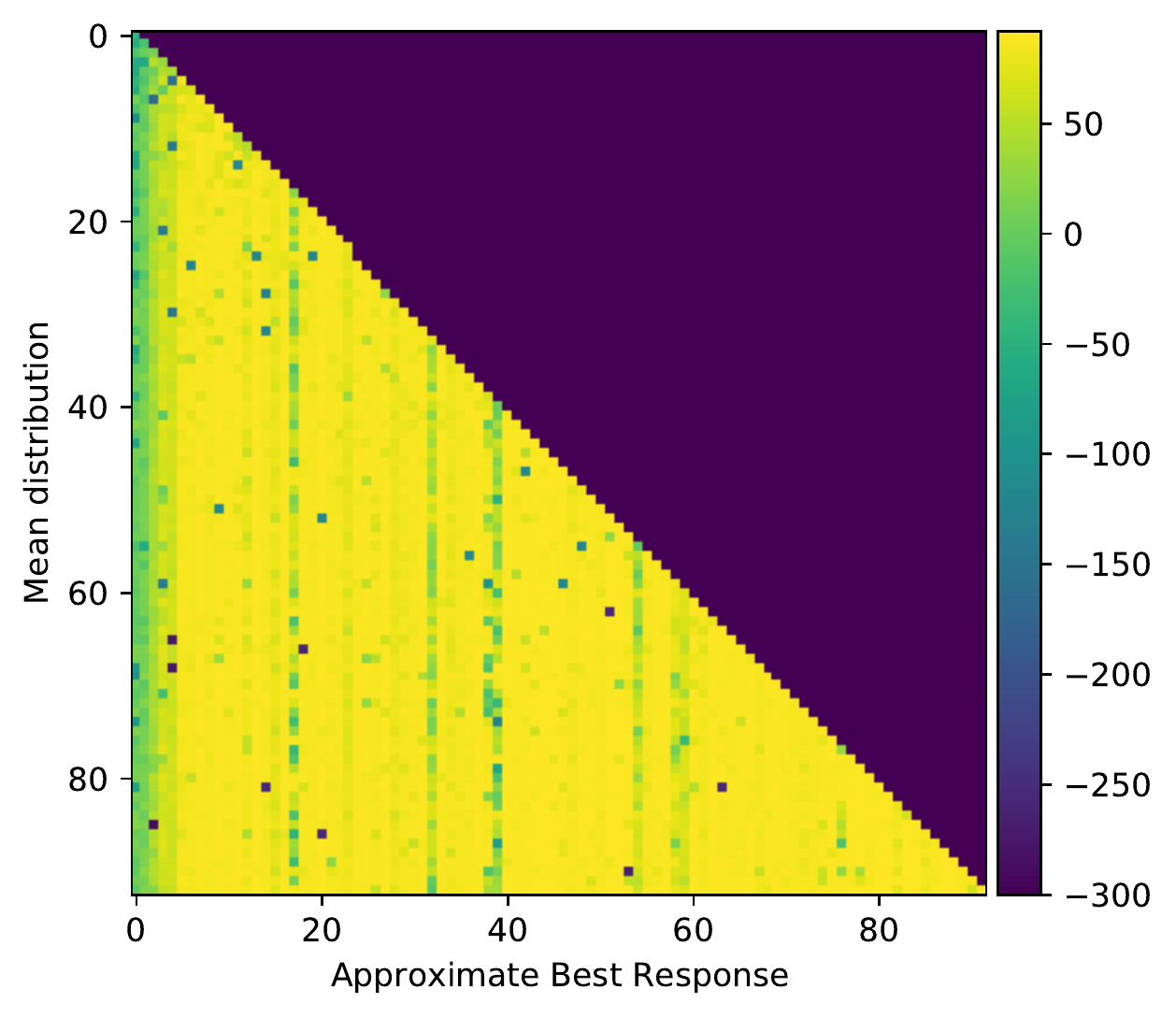}
      \caption{Approximate exploitability}
    \end{subfigure}

    \caption{Flocking in 4D with many obstacles.}
    \label{fig:4d_many_obstacle}
\end{figure}

\section{Normalizing Flows}

In this section we provide some background on Normalizing Flows, which are an important part of our approach.

\paragraph{Coupling layers}
A coupling layer applies a coupling transform, which maps an input $x$ to an output $y$ by first splitting $x$ into two parts $x=[x_{1:d-1}, x_{d:D}]$, computing parameters $\theta = NN(x_{1:d-1})$ of an arbitrary neural network (in our case a fully connected neural network with 8 hidden units), applying $g$ to the last coordinates $y_i = g_{\theta_i}(x_i)$ where $i \in [d, \dots, D]$ and $g_{\theta_i}$ is an invertible function parameterized by $\theta_i$. Finally, we set $y_{1:d-1} = x_{1:d-1}$. Coupling transforms offer the benefit of having a tractable Jacobian determinant, and they can be inverted exactly in a single pass.

\paragraph{Neural Spline Flows}
NSFs are  based on monotonic rational-quadratic splines. These splines are used to model the function $g_{\theta_i}$. A rational-quadratic function takes the form of a quotient of two quadratic polynomials, and a spline uses $K$ different rational-quadratic functions.

Following the implementation described in \cite{durkan2019neural}, we detail how a NSF is computed. 

\begin{enumerate}
    \item A neural network NN takes $x_{1:d-1}$ as inputs and outputs $\theta_i$ of length $3K -1$ for each $i \in [1, \dots, D]$.
    \item $\theta_i$ is partitioned as $\theta_i = [\theta_i^w, \theta_i^h, \theta_i^d]$, of respective of sizes $K, K,$ and $K-1$.
    \item $\theta_i^w$ and $\theta_i^h$ are passed through a softmax and multiplied by $2B$, the outputs are interpreted as the widths and heights of the $K$ bins. Cumulative sums of the $K$ bin widths and heights yields the $K+1$ knots ${(x^{k},y^{k} )}^K_{k=0}$.
    \item $\theta_i^d$ is passed through a softplus function and is interpreted as the values of the derivatives of the internal knots.
\end{enumerate}

\section{Visual Rendering with Unity}

\begin{figure}[htbp]
    
      \begin{subfigure}{0.5\textwidth}
      \centering
      \includegraphics[width=0.9\linewidth]{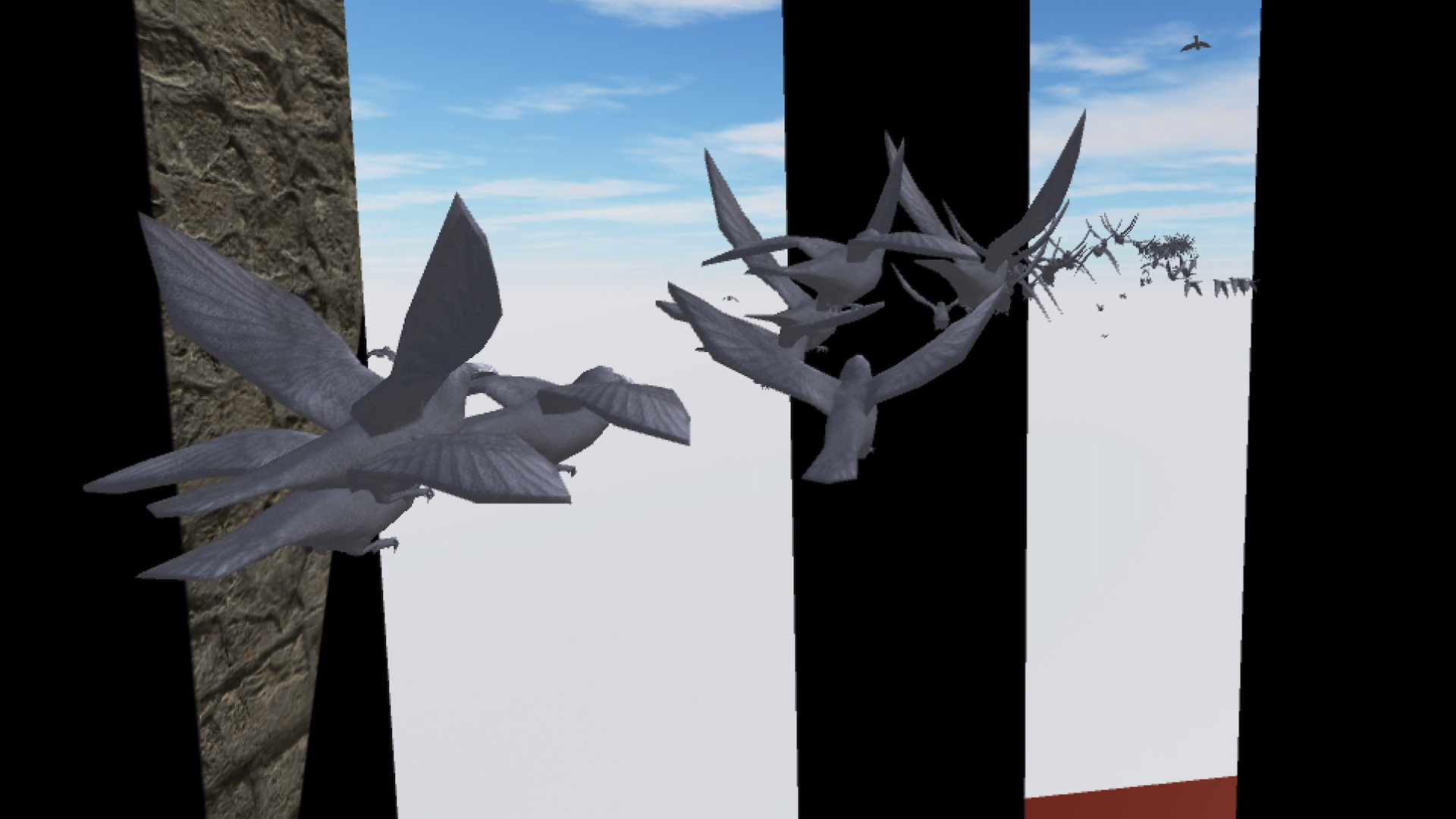}
    \end{subfigure}%
    
    \medskip
    
    \begin{subfigure}{0.5\textwidth}
      \centering
      \includegraphics[width=0.9\linewidth]{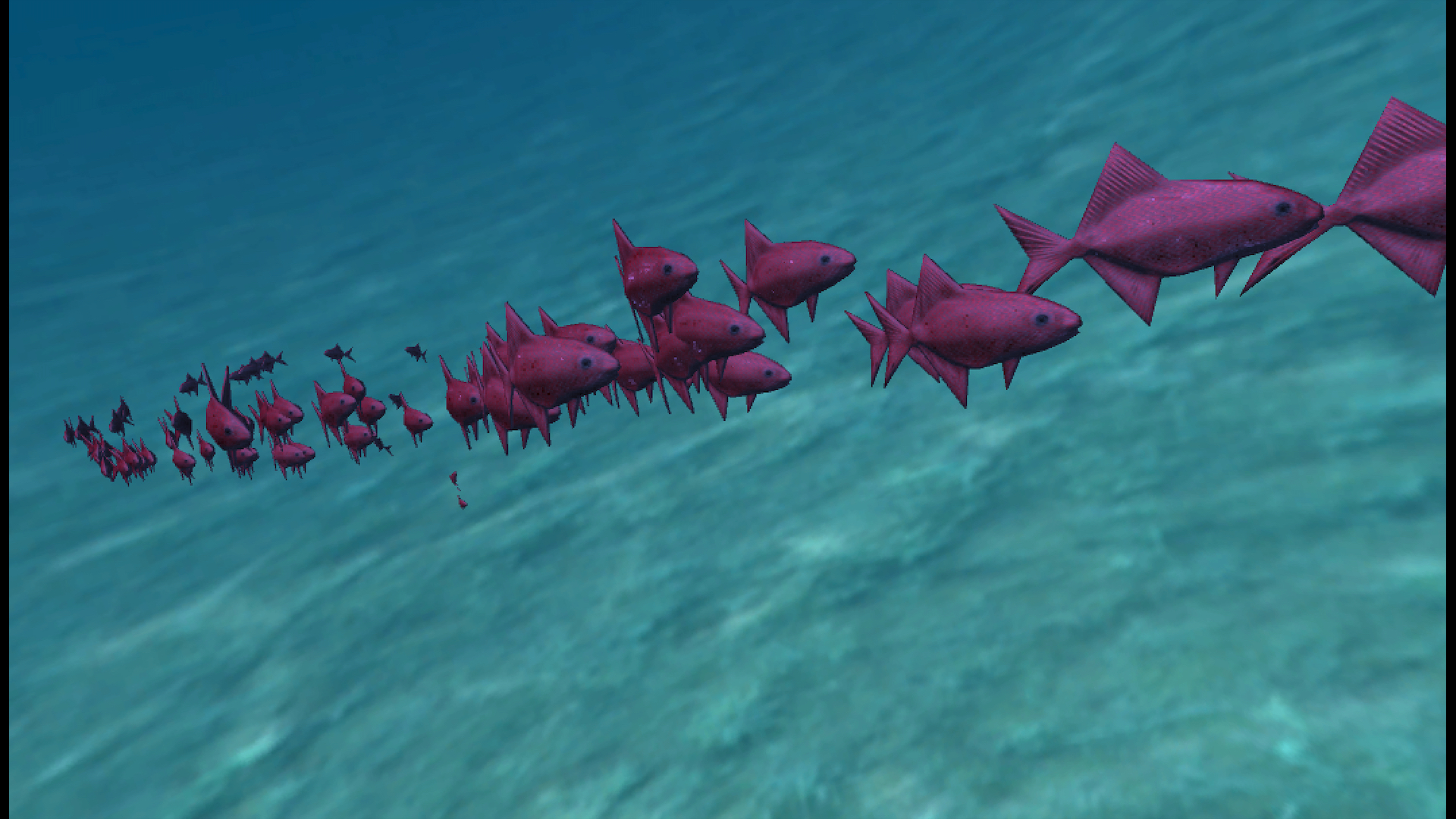}
    \end{subfigure}%

    \caption{Visual rendering with Unity}
    \label{fig:Unity_rendering}
\end{figure}

Once we have trained the policy with the Flock'n RL algorithm, we generate trajectories of many agents and stock them in a csv file. We have coded an integration in Unity, making it possible to load these trajectories and visualize the flock in movement, interacting with its environment. We can then easily load prefab models of fishes, birds, or any animal that we want our agents to be. We can also load the obstacles and assign them any texture we want. Examples of rendering are available in Fig.~\ref{fig:Unity_rendering}.

\end{document}